\documentclass[aps,longbibliography,superscriptaddress,prapplied,notitlepage,reprint]{revtex4-1}
\usepackage{epsfig,color}
\usepackage{graphicx}
\usepackage{dcolumn}
\usepackage{bm}
\usepackage{amsmath,amsfonts,amssymb,mathrsfs}
\usepackage{pstricks}
\usepackage{amsxtra}
\usepackage{amsthm}
\usepackage{braket}
\usepackage{natbib}
\usepackage{physics}
\usepackage{hyperref,color}
\definecolor{darkblue}{rgb}{0.0,0.0,0.7}
\hypersetup{colorlinks,breaklinks,linkcolor=darkblue,urlcolor=darkblue,anchorcolor=darkblue,citecolor=darkblue}

\providecommand{\kb}[1]{\ket{#1}\bra{#1}} 
\providecommand{\ah}{{\hat{a}}}
\providecommand{\ad}{{\hat{a}^\dagger}}
\providecommand{\shz}{{\hat{\sigma}_z}}
\providecommand{\shp}{{\hat{\sigma}_+}}
\providecommand{\shm}{{\hat{\sigma}_-}}
\providecommand{\bh}{{\hat{b}}}
\providecommand{\bd}{{\hat{b}^\dagger}}
\providecommand{\Hh}{{\hat{H}}}
\providecommand{\Hhe}[1]{{\hat{H}_{\rm eff}^{(#1)}}}
\providecommand{\Gh}{{\hat{G}}}
\providecommand{\Vh}{{\hat{V}}}
\providecommand{\Uh}{{\hat{U}}}
\providecommand{\GhI}[1]{\hat{G}^{(#1)}_I} 
\providecommand{\hd}{{\hat{d}}}
\providecommand{\dhd}{{\hat{d}^\dagger}}

\begin{document}
\title{Photon-Number-Dependent Hamiltonian Engineering for Cavities}

\author{Chiao-Hsuan Wang}
\email{chiao@uchicago.edu}
\affiliation{Pritzker School of Molecular Engineering, University of Chicago, Chicago, Illinois 60637, USA}

\author{Kyungjoo Noh}
\affiliation{AWS Center for Quantum Computing, Pasadena, California 91125, USA}

\author{Jos\'{e} Lebreuilly}
\affiliation{Department of Physics and  Yale Quantum Institute, Yale University, New Haven, Connecticut 06520, USA}

\author{S. M. Girvin}
\affiliation{Department of Physics and Yale Quantum Institute, Yale University, New Haven, Connecticut 06520, USA}

\author{Liang Jiang}
\affiliation{Pritzker School of Molecular Engineering, University of Chicago, Chicago, Illinois 60637, USA}

\begin{abstract}
Cavity resonators are promising resources for quantum technology, while native nonlinear interactions for cavities are typically too weak to provide the level of quantum control required to deliver complex targeted operations.  Here we investigate a scheme to engineer a target Hamiltonian for photonic cavities using ancilla qubits.   By off-resonantly driving dispersively coupled ancilla qubits, we develop an optimized approach to engineering an arbitrary photon-number-dependent (PND) Hamiltonian for the cavities while minimizing the operation errors.  The engineered Hamiltonian admits various applications including canceling unwanted cavity self-Kerr interactions, creating higher-order nonlinearities for quantum simulations, and designing quantum gates resilient to noise.  Our scheme can be implemented with coupled microwave cavities and transmon qubits in superconducting circuit systems.
\end{abstract}
\maketitle

\section{INTRODUCTION}
Microwave cavity resonators are rising as a promising platform for quantum information processing.  Tremendous experimental progress has been made in building high-coherence microwave photon cavities in circuit quantum electrodynamics (cQED) platforms~\cite{Schoelkopf2008,Devoret2013,Kjaergaard2019,Lei2020}.  The infinite-dimensional Hilbert space of a single resonator enables flexible and hardware-efficient design of quantum error correction codes~\cite{Gottesman2001,Mirrahimi2014,Michael2016,Albert2018,Grimsmo2019,Grimm2020} and has led to the success in extending the logical qubit lifetime \cite{Ofek2016}.  Controllable cavity systems can also be used to emulate the dynamics of the classically intractable many-body quantum systems due to their rapidly growing Hilbert space~\cite{Hartmann2016,Noh2017}. Recent success in realization of boson sampling of microwave photons to emulate the optical vibrational spectra of triatomic molecules \cite{Wang2020} is an example of an early experimental step towards this goal.

 The advantages brought by the flexible Hilbert space structure of cavity resonators are accompanied by crucial challenges to manipulate such systems.   General quantum operations across several photon-number states require highly nonlinear interactions, which are also crucial for many-body photonic quantum simulations. However, the native nonlinear interactions among photons are often weak and untunable. On the other hand, Hamiltonian engineering utilizes controlled operations to generate tailored evolution to deliver complicated tasks beyond the capacity of native interactions, that can be applied to quantum information processing, quantum sensing, and quantum simulation~\cite{Schirmer2006,Goldman2014,Krantz2019,Haas2019}. Inspired by advances in the universal control of microwave cavity modes using an ancilla superconducting qubit~\cite{Krastanov2015,Heeres2015,Heeres2017,Gao2019}, here we develop a general formalism to engineer a photon-number-dependent (PND) Hamiltonian for cavities appropriate for cQED devices.

In Sec.~\ref{sec:dispersive}, we study the time evolution of a dispersively coupled qubit-cavity system under off-resonant drives. We then propose in Sec.~\ref{sec:PND} a general protocol to design optimized drives that can engineer a target PND Hamiltonian for a single cavity, and discuss cavity dephasing induced by the ancilla qubit decohernece in Sec.~\ref{sec:decoherence}. We further extend our method to include higher-order corrections to the system Hamiltonian in Sec.~\ref{sec:Kerr}, and to implement a fault-tolerant gate between coupled cavities in Sec. ~\ref{sec:coupled}. We conclude in Sec.~\ref{sec:summary} by summarizing our results and motivating potential quantum computation and quantum simulation applications.

\section{DISPERSIVE MODEL WITH OFF-RESONANT DRIVES \label{sec:dispersive}}
We first consider a dispersively coupled~\cite{Boissonneault2009} qubit-cavity system described by the Hamiltonian
\begin{align}
\Hh_0= \hbar \omega_a \ad \ah +  \hbar\omega_q \kb{e}- \hbar \chi \ad \ah\kb{e},
\label{eqn:H0}
\end{align}
where $\omega_a$ is the frequency of the cavity mode $\ah$, $\omega_q$ is the qubit transition frequency between qubit states $\ket{g}$ and $\ket{e}$, and $\chi$ is the dispersive coupling strength.  The effective qubit transition frequency is dependent upon the number states of the cavity, $\ket{n}$, with resonant frequencies $\omega_{q,n}=\omega_q-\chi n $.  Such dispersive interaction between superconducting qubits and microwave cavities has been a useful resource for quantum control and readout in cQED devices~\cite{Schuster2007,Boissonneault2009,Heeres2015,Heeres2017,Gao2019}.

\begin{figure}[htbp]
\begin{center}
\includegraphics[width=.45 \textwidth]{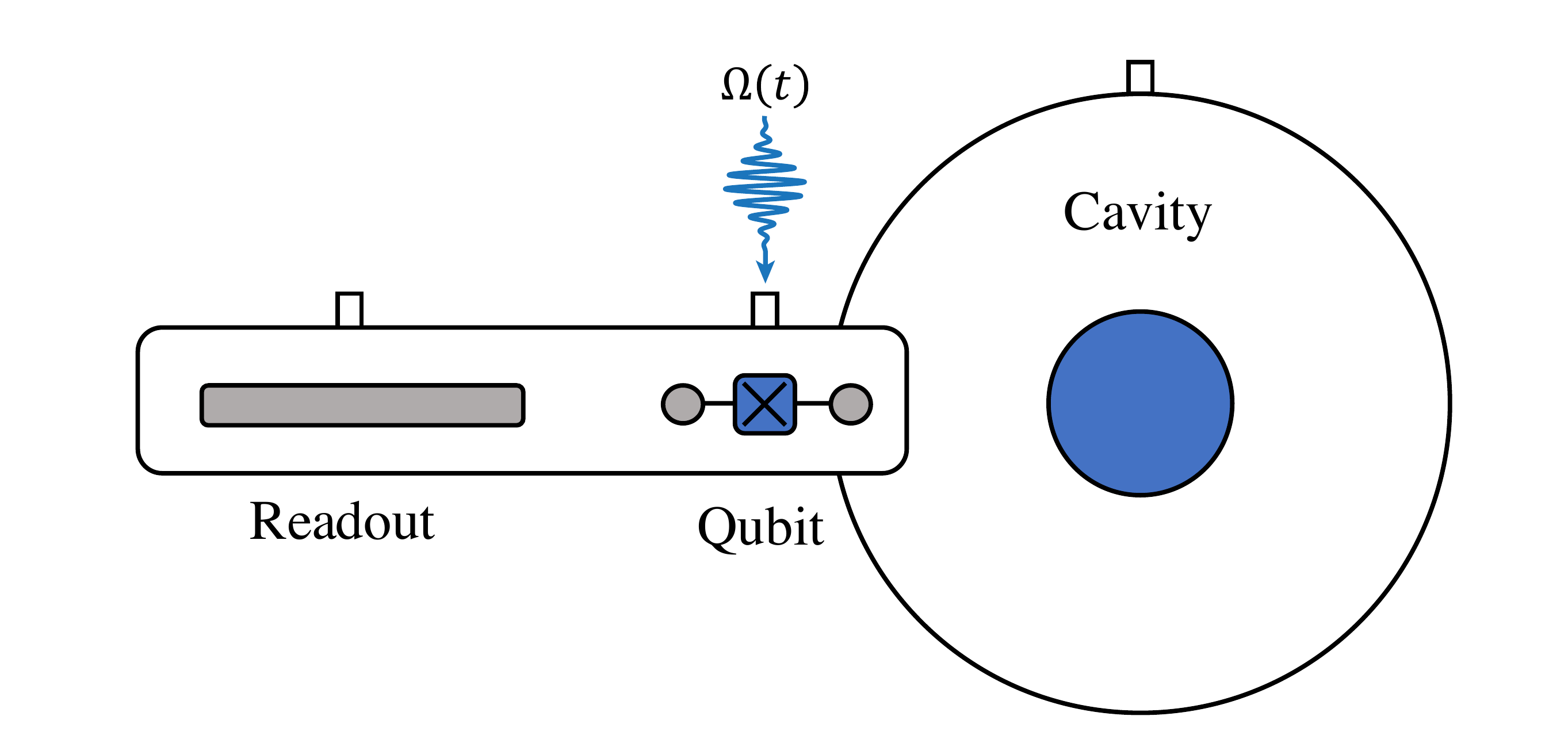}
\caption{Schematic of the cQED implementation of a coupled qubit-cavity system, where a transmon qubit is capacitively coupled to a three-dimensional (3D) microwave cavity resonator. One can apply control drives on the transmon qubit through a drive port, and the qubit can be coupled to an additional quasiplanar resonator for readout.}
\label{fig:cqed} 
\end{center}
\end{figure}

Applying a time-dependent drive $\Omega(t)$ to the qubit,
\begin{align}
\Vh(t)= \hbar \Omega(t) \shm +\hbar \Omega^*(t) \shp,
\label{eqn:H1}
\end{align}
and working in the number-split regime~ \cite{Gambetta2006,Schuster2007,Johnson2010}, where $\chi$ is larger than the transition linewidth of both the qubit and the cavity, one can drive the qubit near selective number-dependent transition frequencies to address individual number states of the cavity (see schematic diagram in Fig.~\ref{fig:cqed}).  In contrast to the recently demonstrated scheme of imparting selective number-dependent arbitrary phases (SNAP) to photon Fock states by directly exciting qubit transitions \cite{Krastanov2015,Heeres2015}, here we work in the large drive detuning regime to engineer a continuous photon-number-dependent target Hamiltonian.

\subsection{Unitary evolution with abrupt drives}
We consider control drives of the form $\Omega(t)=\sum_{m \in \mathbb{Z}}  \Omega_m e^{i ( \omega_q- {m} \chi +\delta_{m} )t}$.  Moving the total Hamiltonian $\Hh(t)=\Hh_0+\Vh(t)$ to the interaction picture with the unitary transformation $\Uh=\exp(i \Hh_0 t /\hbar)$, we obtain
\begin{align}
\Vh_I(t)=\sum_m\sum_n \hbar \Omega_m (e^{i [(n-m) \chi +\delta_m] t} \kb{n} \shm + \mathrm{H.c.}).
\end{align}
For this periodically driven qubit-cavity system, we assume a tripartition ansatz for the evolution operator $\ket{\psi(t_f)}_{I}$~\cite{Rahav2003} , 
\begin{align}
\Uh_{I}(t_f,t_i)=e^{-i \Gh_I(t_f)}e^{-i\Hh_{\rm eff,I}(t_f-t_i)/\hbar}e^{i \Gh_I(t_i)}, 
\end{align}
where the subscript $I$ denotes evolution in the interaction picture.

Assuming
$\forall m, \abs{\Omega_{m}} \ll \abs{\delta_{m}},\abs{\chi -\delta_m}$, we use time-dependent perturbation theory to find an effective Hamiltonian, 
\begin{align}
\Hh_{\rm I,eff} =&-\sum_m\sum_n\frac{\hbar|\Omega_m|^2 \kb{n}\shz}{(n-m)\chi+\delta_m}
+ \mathcal{O}(\Vh^4),
\label{eqn:Heff}
\end{align}
which governs the long time dynamics of the system up to the initial and final kicks, $\Gh_I(t_i)$ and $\Gh_I(t_f)$(see Appendix~\ref{A:Dispersive} for detailed derivations).  Since we are only driving the qubit off-resonantly with $\forall m, \abs{\Omega_{m}} \ll  \abs{\delta_{m}}, \abs{\chi -\delta_m} $, we can assume that it stays in its ground state.  Moving back to the original frame, the effective Hamiltonian seen by the photon while the qubit stays in its ground state is
\begin{align}
\Hh_{\rm  eff,g} =\hbar \omega_a \ad \ah+\bra{g} \Hh_{\rm I, eff} \ket{g} =\hbar \omega_a \ad \ah+ \Hh_E.
\label{Heffg}
\end{align}
The off-resonant control drives on the ancilla qubit thus effectively generate a photon-number-dependent Hamiltonian $\Hh_E=\sum_n \hbar E_n \kb{n}$ for the cavity .

Rapidly oscillating micromotion is predicted by the kick operator~\cite{Rahav2003,Goldman2014}.  The leading-order kick operator is 
\begin{align}
\GhI{1}(t)=\sum_n \sum_m \frac{\Omega_m \kb{n}\left(\shm e^{i((n-m)\chi+\delta_m )t}-\text{H.c.} \right)}{i[(n-m)\chi+\delta_m]}.
\label{G1}
\end{align}
To the first order in $\Vh$, an initial state $\ket{n,g}$ will evolve to $\ket{n,g}+\sum_m \frac{\Omega_m \kb{n}}{(n-m)\chi+\delta_m}(e^{-i[(n-m)\chi+\delta_m ]t} -1)\ket{n,e}$ at time $t$, showing an oscillating small population of the qubit excited state component $\ket{n,e}$ with a time-averaged probability $p_{n,e}=\sum_m |\frac{\Omega_m}{(n-m)\chi+\delta_m}|^2+|\sum_m \frac{\Omega_m}{(n-m)\chi+\delta_m}|^2$, where the second term is the contribution from the initial kick at $t=0$.  This excited state component can be viewed as coherent oscillations assuming a closed qubit-cavity system.  If one chooses detunings commensurate with the dispersive coupling strength $\chi$, the overall micromotion vanishes at a period $T_{M}=2 \pi/  \text{GCD}(\left\{\delta_m{'}s, \chi\right\})$, where $\text{GCD}(\left\{\delta_m{'}s, \chi\right\})$ is the greatest common divisor among all the detunings and the dispersive shift,  and averages to zero at long time.  For quantum gates implemented by PND Hamiltonian, it is essential to design drives such that $T_G = c T_M$ for some $c \in \mathbb{N}$ in order to achieve maximum gate fidelity.  Alternatively, one can relax this constraint on $T_G$ by smoothly turning on and off the drive to remove the effect of the initial and the final kicks.

\subsection{Unitary evolution with smooth ramping}
So far we assume that the drive is abruptly turned on at an initial time $t_i$ and lasts till a final time $t_f$.  One can alternatively apply a ramping function $\lambda(t)$ such that $\Hh(t)=\Hh_0 + \lambda(t)\Vh(t)$ to smoothly turn on (and off) the drive, which will remove the effect associated with the initial (and the final) kick operator if the ramping time scale is much longer than $1/\chi$.  The choice of the ramping function $\lambda(t)$ is not unique. 

For mathematical simplicity, we first consider the case of applying a sinusoidal envelope $\lambda(t)=\sin(\gamma t)$ to a short-time gate operation from $t=0$ to $t=T_G=\pi/\gamma$. Using the time-dependent perturbation theory, we find
\begin{align}
&-\frac{i}{\hbar} \int_{0}^{T_G}\sin(\gamma t_1) \Vh_I(t_1)dt_1
\xrightarrow{\chi \gg \gamma} 0,
\end{align}
and
\begin{align}
\left(\frac{-i}{\hbar}\right)^2& \int_{0}^{T_G}dt_1\int_{0}^{t_1}dt_2\sin(\gamma t_1)\Vh_I(t_1)\sin(\gamma t_2)\Vh_I(t_2) \notag\\&\xrightarrow{\chi \gg \gamma} \sum_m \sum_n \frac{i |\Omega_m|^2 \kb{n}}{2[(n-m)\chi+\delta_m]}\shz T_G.
\end{align}

In the limit $\chi \gg \gamma$, the resulting time evolution with this smooth sinusoidal envelope is thus equivalent to having an effective Hamiltonian generated by $\Vh(t)/\sqrt{2}$ but without any initial or final kick effects. To compensate for the $1/\sqrt{2}$ factor, one can implement the same gate (by accumulating the same phase) as the abrupt version $\Hh(t)=\Hh_0+\Vh(t)$ by rescaling the ramping function to $\lambda_{\rm gate}(t)=\sqrt{2}\sin(\gamma t)$ or by letting the system evolve twice as long.  Note that in the abrupt version one has to choose a gate time at which the micromotion vanishes, while with the sinusoidal envelope there is no such requirement because the micromotion has already been removed by the smooth ramping.

For long-time operation of the PND Hamiltonian engineering scheme, one can design a ramp-up function $\lambda_{\rm up}(t)= 0 \rightarrow 1 $ and a ramp-down function $\lambda_{\rm down} (t)= 1 \rightarrow 0$ at the beginning and the end of the drive.  Here we provide one example of the ramp-up and ramp-down functions,
\begin{widetext}
\begin{align}
 \lambda_{\rm up}(t) =
    \begin{cases}
      \lambda_s \sin[\frac{\pi (t-t_i)}{2T_s}]  & t_i \leq t \leq t_i+T_s\\
      \frac{\lambda_s-1}{2} \sin[\frac{\pi (t-t_i)}{2T_s}]+\frac{\lambda_s+1}{2} & t_i+T_s \leq t \leq t_i+3 T_s
    \end{cases},
\label{eqn:lambdaup}
\end{align}
\begin{align}
 \lambda_{\rm down}(t)=
    \begin{cases}
      \frac{\lambda_s-1}{2} \sin[\frac{\pi (t_f-t)}{2T_s}]+\frac{\lambda_s+1}{2} & t_f - 3 T_s \leq t \leq t_f - T_s\\
      \lambda_s \sin[\frac{\pi (t_f-t)}{2T_s}] &t_f - T_s \leq t \leq t_f
    \end{cases},
\label{eqn:lambdadown}
\end{align}
\end{widetext}
and $\lambda(t)=1$ otherwise.  Here $\lambda_s=\frac{\sqrt{46}-1}{5}$ is a special chosen value to guarantee the same accumulated phase as the abrupt case during the ramp-up and ramp-down periods.

\section{PND HAMILTONIAN ENGINEERING FOR A SINGLE CAVITY \label{sec:PND}}
Given a target Hamiltonian,
\begin{align}
\Hh_{\rm  T} =\sum_n \hbar E_{T,n} \kb{n},
\end{align}
one may find appropriate values of $\Omega_{m}$ and $\delta_{m}$ such that $\Hh_E=\Hh_T$. The solution for $\Omega_{m}$ and $\delta_{m}$ for a given target Hamiltonian (with reasonable strengths $E_{T,n} \ll \chi$) is not unique.  Here we suggest a way of designing the drives as described below.

First, we consider a finite set of possible detunings $\delta_m{'}s =\left\{ \pm \chi/2,\pm \chi/4 \right\}$. By selecting detunings commensurate with $\chi$, we can ensure that there are no suprising near-resonant higher-order contributions and also easily determine the periodicity at which the micromotion vanishes, $T_M=8 \pi/\chi$ for the chosen set of $\delta_m{'}s$ (or $T_M=4 \pi/\chi$ if $\delta_m{'}s =\left\{ \pm \chi/2\right\}$).  Those detunings are comparable to $\chi$ which allows the largest possible engineered Hamiltonian strength.  Second, we assign random choices of drive detunings from $\delta_m{'}s$ for each number state and find the optimized parameters that generate the target Hamiltonian according to Eq.~(\ref{eqn:Heff}) plus fourth-order perturbation theory terms while minimizing $\sum_n p_{n,e}$, the summation of the average qubit excited-state probability due to micromotion. The optimized choice also minimizes the decoherence induced by qubit relaxation, which is discussed later.

Below we present concrete examples to demonstrate versatile applications of PND, with numerical simulation results shown in Fig.~\ref{fig:simulations} (optimized parameters displayed in Appendix~\ref{A:parameters}).  Assuming a dispersive shift $\chi/2 \pi=2.56$ MHz appropriate for coupling between transmon qubit and superconducting cavity resonator in  cQED devices~\cite{Heeres2017}, we are able to engineer Hamiltonian strengths up to $E_T/2\pi \approx 50$ kHz with high precision. Even larger strengths $E_T/2\pi \approx 150$ kHz are achievable but are subject to imperfections due to sixth- and higher-order terms in the perturbation theory.  This energy scale of the engineered Hamiltonian, $E_T /2\pi \approx 50$ kHz (150 kHz), is much larger than the cavity decoherence rate $\kappa_a /2\pi \approx 0.01$ kHz for state-of-the-art 3D microwave cavities~\cite{Reagor2013} and is thus favorable to achieve high-fidelity gates or to perform quantum simulation.

\subsection{Photon-photon interaction}
One direct application for PND Hamiltonian engineering is to create tunable photon-photon nonlinear interactions to emulate dynamics of quantum many-body systems with cavity photons~\cite{Hartmann2016,Noh2017}. Such nonlinearities are typically weak in native interactions.  For example, one can engineer a purely three-photon interaction for cavity photons by setting
\begin{align}
\Hh_T=\Hh_3 =\sum_n  \hbar K_3 n(n-1)(n-2) \kb{n}.
\end{align}

\subsection{Parity-dependent energy}
Photon-number parity serves as an error syndrome in various bosonic quantum error correction codes such as cat codes and binomial codes~\cite{Mirrahimi2014,Michael2016}. By engineering a Hamiltonian of the form
\begin{align}
\Hh_T=\Hh_P =\sum_n  \hbar P (-1)^{n+1} \kb{n},
\end{align}
 the cavity can distinguish photon-number parity by energy, which might allow us to design error detection or dynamical stabilization of the code states for bosonic quantum error correction~\cite{Lebreuilly2021}.
 
\subsection{Error-transparent Z-rotation}
Continuous rotation of the encoded logical qubit around the Z axis can generate the whole family of phase-shift gates $R_{\theta}$, including $\pi/8$ gate and Z gate, which are common elements of single-qubit gates for universal quantum computing~\cite{Nielsen2010}. For quantum information encoded in rotational-symmetric bosonic code that can correct up to $(d_n-1)$-photon loss errors~\cite{Grimsmo2019}, the logical states are
\begin{align}
\ket{0_{d_n}}_L\equiv\sum_{k=0}^{\infty} f_{2k d_n} \ket{n=2 k d_n},
\end{align}
\begin{align}
\ket{1_{d_n}}_L\equiv\sum_{k=0}^{\infty} f_{(2k+1) d_n} \ket{n=(2 k+1) d_n},
\end{align}
with code-dependent coefficients $f_n${'}s. Phase-shift gates at an angle $\theta$ for logical states can be implemented via the cavity Kerr effect $\propto(\ad\ah)^2$ for the Z gate $\theta=\pi/2$ ~\cite{Mirrahimi2014,Grimsmo2019} or by four-photon interaction $\propto (\ad\ah)^4$ for the $\pi/8$-gate $\theta=\pi/4$~\cite{Grimsmo2019}.

To achieve fault-tolerant quantum computation, one can instead design an error-transparent~\cite{Vy2013,Kapit2018,Rosenblum2018,Sun2020} Hamiltonian, that commutes with and is thus uninterrupted by the photon-loss error, to perform continuous logical Z rotations.  By engineering the same positive energy shift $\hbar g_R$ for $\ket{0}_L$ and all of its recoverable error states while engineering an equal but opposite energy shift $-\hbar g_R$ for $\ket{1}_L$ and all of its recoverable error states, the resulting Z rotation is `transparent' to $(d_n-1)$-photon-loss errors.  Specifically, for cat codes or binomial codes with $d_n=2$,
\begin{align}
\Hh_Z &=\sum_{k=0}^{\infty}  \hbar g_R (\kb{4k}+\kb{4k+3}\notag\\&-\kb{4k+2}-\kb{4k+1}).
\end{align}

\begin{figure}[htbp]
\begin{center}
\includegraphics[width=0.45 \textwidth]{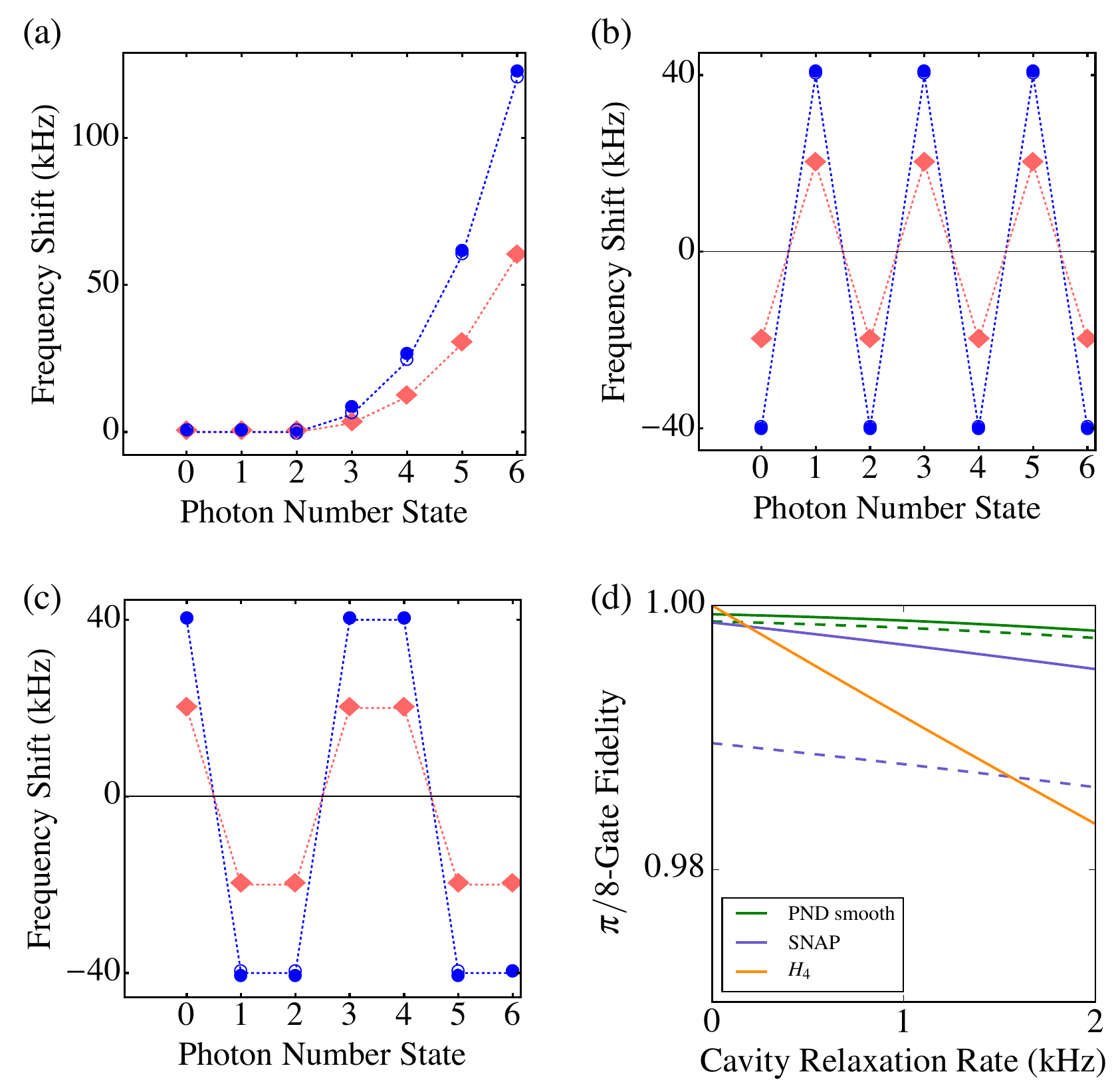}
\caption{Simulation of the PND Hamiltonian engineering frequency shifts for (a) three-photon interaction, (b) parity-dependent energy, and (c) error-transparent Z rotation.  The target Hamiltonian ($E_{T,n}/2\pi$) is shown in dotted lines with open markers, the engineered Hamiltonian ($E_{n}/2\pi$) is represented by solid markers, and different colors represent different engineered Hamiltonian strengths. (d) Fidelity of $\pi/8$ gate as a function of the cavity relaxation rate.  Dashed lines represent the modifications due to ancilla qubit relaxation at a rate $\Gamma_q/2\pi=3$ kHz. Here we assume $\chi/2\pi=2.56$ MHz, $\pi/8$-gate time $T_G=16\pi/\chi=2 T_M$, and a smooth PND ramping function $\lambda_{\rm gate}(t)=\sqrt{2}\sin(\pi t /T_G)$. The PND parameters are displayed in Tables~\ref{tab:threephoton1}-\ref{tab:piovereight2}.}
\label{fig:simulations} 
\end{center}
\end{figure}

Consider the $\pi/8$ gate ($\theta=\pi/4$) on the kitten code $\ket{0}_k=(1/\sqrt{2})(\ket{0}+\ket{4})$,  $\ket{1}_k=\ket{2}$~\cite{Michael2016} for example. This rotation can be implemented by applying $\Hh_Z$ for a time $t=\pi/8 g_R$, by imparting phase $-\pi/8$ on $\ket{n=0,3,4}$ and phase $+\pi/8$ on $\ket{n=1,2}$ with a SNAP gate, or by applying $H_4=\hbar K_4 (\ad\ah)^4 $ for a time $t=\pi/64 K_4$. 
We characterize the gate performance in the presence of photon-loss by performing the rotation gate on $(1/\sqrt{2})(\ket{0}+\ket{4}) \otimes \ket{g}$ over the same gate time, followed by instantaneous recovery of single-photon loss error~\cite{Michael2016,Lihm2018} in Fig.~\ref{fig:simulations}(d).  Comparing the final fidelities, the PND gate and the SNAP gate show much higher resilience to photon-loss error than $H_4$ due to their error-transparent structure.

\section{Qubit-induced decoherence \label{sec:decoherence}}
In practice, the decoherence of the qubit may induce cavity dephasing during the PND process.  Specifically, the qubit relaxation jump operator $\shm$ at a rate $\Gamma_q \ll \chi$ would cause dephasing for off-diagonal density matrix elements of the cavity number states $\rho_{n_1 n_2}$ at a rate $\gamma_{n_1 n_2}=\Gamma_q/2 (p_{n_1,e}+p_{n_2,e})$, while the qubit dephasing jump operator $\kb{e}$ at a rate $\Gamma_{\phi} \ll \chi$ causes cavity dephasing at a rate $\gamma_{n_1 n_2}=(\Gamma_{\phi}/2) \{p_{n_1,e}+p_{n_2,e}-2\sum_{m_1}  \Omega_{m_1}^{*}/[(n_1-m_1)\chi+\delta_{m_1}]\sum_{m_2} \Omega_{m_2}/[(n_2-m_2)\chi+\delta_{m_2}]\}$(see Appendix~\ref{A:Decoherence}). Our choice of the optimized parameters for minimizing the micromotion also minimizes the decoherence induced by qubit relaxation, which is the dominant source of imperfection in typical cQED devices with a kHz-order $\Gamma_q$.

\begin{figure}[htbp]
\begin{center}
\includegraphics[width=0.45 \textwidth]{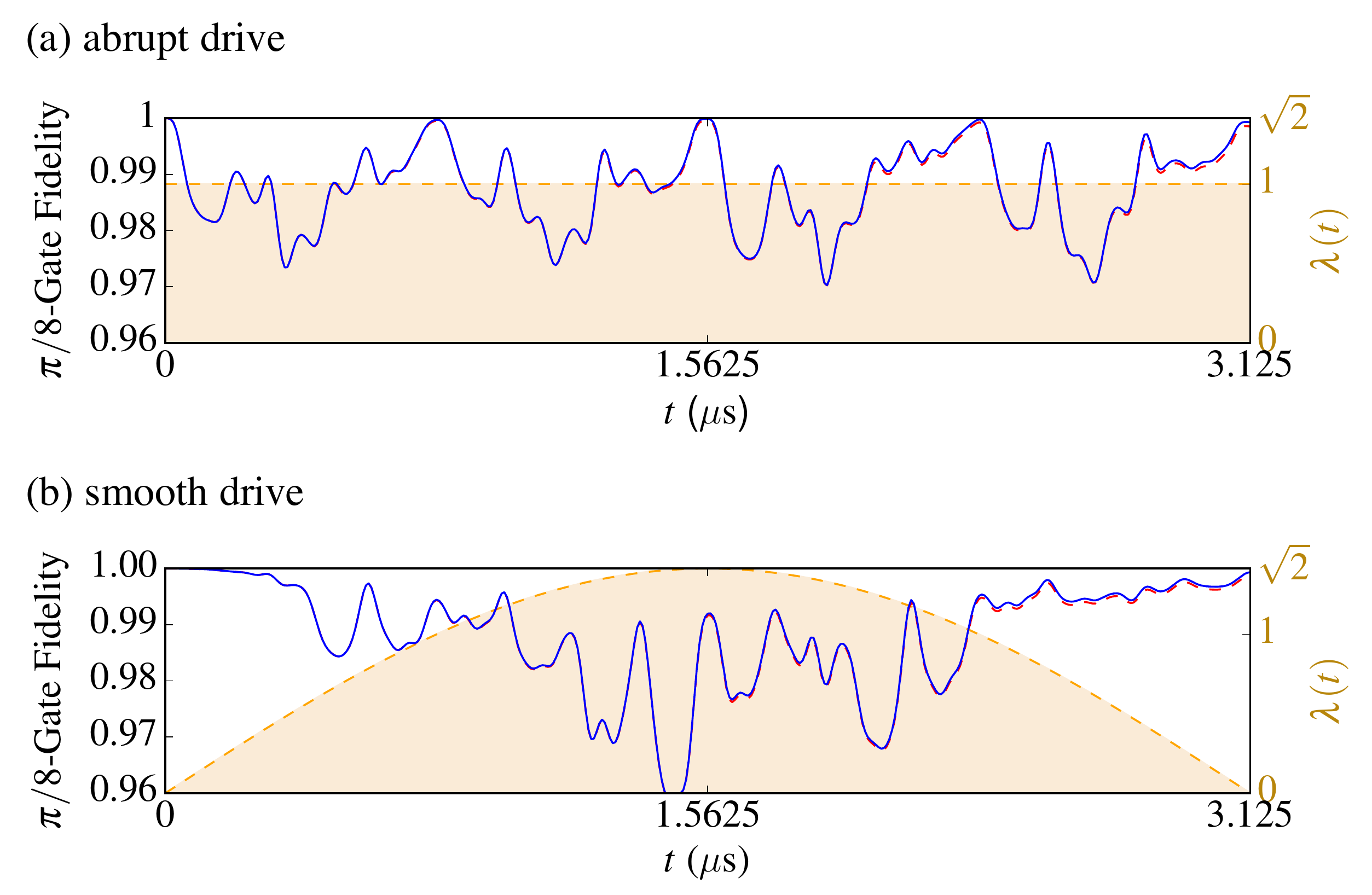}
\caption{Fidelity during the $\pi/8$-gate operation by (a) abruptly turning on the PND drive with $\lambda_{\rm abrupt}(t)=1$ and (b) smoothly turning on and off the PND drive with $\lambda_{\rm gate}(t)=\sqrt{2}\sin(\pi t /T_G)$. The simulated fidelity between the states evolved by the PND drive and the states evolved by the target Hamiltonian $H_Z$ as a function of time is shown in blue (without ancilla qubit relaxation) and red (with ancilla qubit relaxation) curves. The ramping function $\lambda(t)$ in presented as the filled orange curve.  Here we use the parameters in table ~\ref{tab:piovereight} and assume $\Gamma_q/2\pi=3$ kHz and $T_G=16 \pi/\chi=3.125$ $\mu$s.   The initial state is  $(1/\sqrt{2})(\ket{0}_k+\ket{1}_k) \otimes \ket{g}$ .}
\label{fig:smoothgate} 
\end{center}
\end{figure}

Smoothly turning on the PND drive will remove the contribution to $p_{n,e}$ from the initial kick and further reduce the cavity dephasing. In Fig.~\ref{fig:smoothgate} we compare the $\pi/8$-gate operation via the abrupt PND drive versus the smooth PND drive.  At the end of the gate operation, the simulated final gate fidelity is 99.929\% for the abrupt drive and 99.934\% for the smooth drive. The additional infidelity induced by ancilla relaxation is 0.075\% for the abrupt drive and 0.055\% for the smooth drive, showing a reduction in the qubit-induced cavity dephasing by using smooth ramping.

\begin{figure}[htbp]
\begin{center}
\includegraphics[width=0.45 \textwidth]{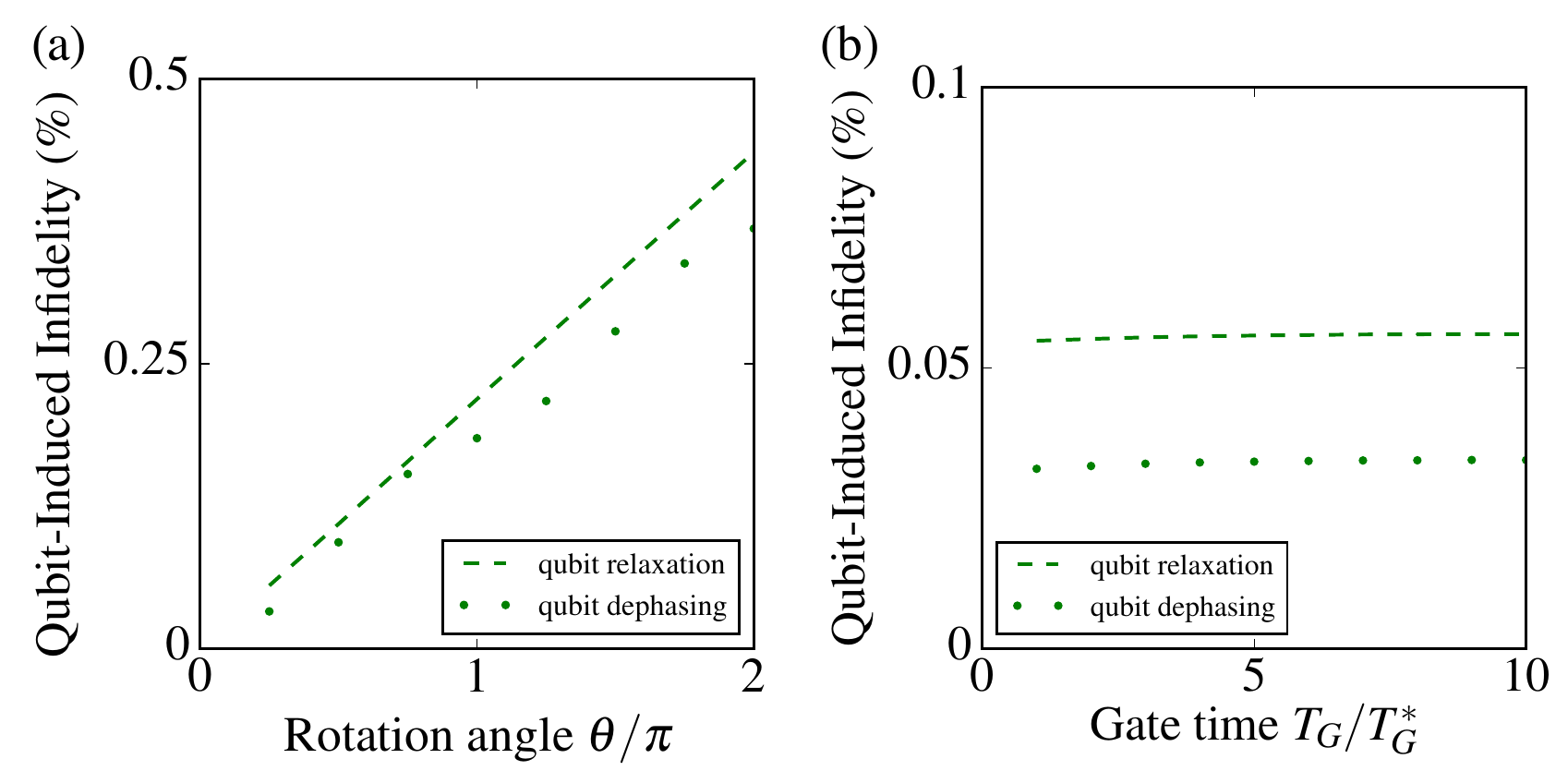}
\caption{Qubit-induced infidelity of the PND $R_{\theta}$ gate. (a) Qubit-induced infidelity as a function of the rotation angle $\theta$, with fixed PND drive parameters $\chi_0, \{ \Omega_0  \}$, $\{ \delta_0  \}$ and with a $\theta$-dependent gate time $T_G= (4\theta/\pi) T_G^{*}$. (b) Qubit-induced infidelity as a function of the gate time $T_G$, with fixed parameters $\theta=\pi/4$, $\chi_0$, $\{ \delta_0  \}$, and with $T_G$-dependent drive amplitudes $\{ \Omega \}=\{ \Omega_0 \sqrt{T_G^{*}/T_G}  \}$. Here $\chi_0, \{ \Omega_0  \}$, and  $\{ \delta_0  \}$ are the parameters in table ~\ref{tab:piovereight}. We assume $\Gamma_q/2\pi=\Gamma_{\phi}/2\pi=3$ kHz, and $T^{*}_G=16 \pi/\chi=3.125$ $\mu$s. The initial state is  $(1/\sqrt{2})(\ket{0}_k+\ket{1}_k) \otimes \ket{g}$ , and we assume a smooth ramping function $\lambda_{\rm gate}(t)=\sqrt{2}\sin(\pi t/T_G)$ for the drive.}
\label{fig:PNDscaling} 
\end{center}
\end{figure}

In contrast to the resonantly-driven SNAP gate which has an averaged qubit excited-state probability $1/2$ during the operation, our scheme has a suppressed qubit excitation and thus has a much smaller decoherence rate during the operation. At the end of the gate operation, the overall qubit-induced decoherence for the PND gate scales as $\Gamma_q \Omega_n^2 T_G/2 \chi^2 \approx |\phi_n| \Gamma_q/2 \chi$, where $\phi_n$ is the phase imparted on the number state $\ket{n}$, while the qubit-induced overall decoherence for the SNAP gate scales as $\Gamma_q T_G/2=\pi \Gamma_q/\Omega$ regardless of the phase (limited by $|\Omega| \ll \chi$).  In Fig.~\ref{fig:PNDscaling}, we study the qubit-induced infidelity for $R_{\theta}$ gate implemented by smooth PND drive. The qubit-induced gate infidelity is proportional to the rotation angle $\theta$ (and thus the total phase) while relatively independent of the gate time while $\theta$ is fixed, as predicted.

The SNAP and PND schemes complement each other for photon-number-dependent operations. The SNAP gate is ideal for one-shot operation to impart large phases. On the other hand, the PND Hamiltonian engineering scheme is better suited for quantum simulation, continuous operation, and quantum gate with small phases. In Fig.~\ref{fig:simulations}(d) we show the $\pi/8$-gate fidelity modified by a lossy qubit in dashed lines. The off-resonantly driven PND gate accumulates much less decoherence (qubit-induced infidelity=0.055\%) than the SNAP gate (qubit-induced infidelity=0.91\%), assuming no cavity relaxation.  Since the qubit-induced decoherence  for the PND gate is proportional to the imparted phase, the maximal qubit-induced PND $R_{\theta}$ gate infidelity is 0.44\% for $\theta=2\pi$, which suggests that the PND scheme shall outperform the SNAP scheme (with the given gate time) for arbitrary error-transparent $R_{\theta}$ gate.

\section{PND Hamiltonian engineering for a single cavity with Kerr \label{sec:Kerr}}
So far we work with the dispersive model of the qubit-cavity coupling. In reality, the underlying microscopic model of coupled qubit-cavity system also predicts higher-order coupling terms ~\cite{Koch2007,Kirchmair2013}. Now consider a generalized model with photon self-Kerr $K$ and second-order dispersive shift $\chi{'}$, the Hamiltonian reads
\begin{align}
\Hh_0&= \hbar \omega_a \ad \ah +   \hbar \omega_q \kb{e} -\hbar \chi\ad \ah \kb{e} - \frac{\hbar K}{2} \ad\ad\ah\ah \notag\\&+\frac{\hbar \chi{'}}{2} \ad\ad\ah\ah\kb{e}.
\end{align}
Adding control drives $\Omega(t)=\sum_m  \Omega_m e^{i ( \omega_q- m \chi+ \delta_m)}$ and  assuming
$\forall m, \abs{\Omega_{m}} \ll \abs{\delta_{m}}$, one can again use time-dependent perturbation theory to find an effective Hamiltonian similar to Eq.~(\ref{eqn:Heff}) but with every $n \chi $ replaced by $n \chi - \chi{'} n (n-1)/2$ due to the second-order dispersive shift (see Appendix~\ref{A:Kerr}).

\begin{figure}[htbp]
\begin{center}
\includegraphics[width=0.45 \textwidth]{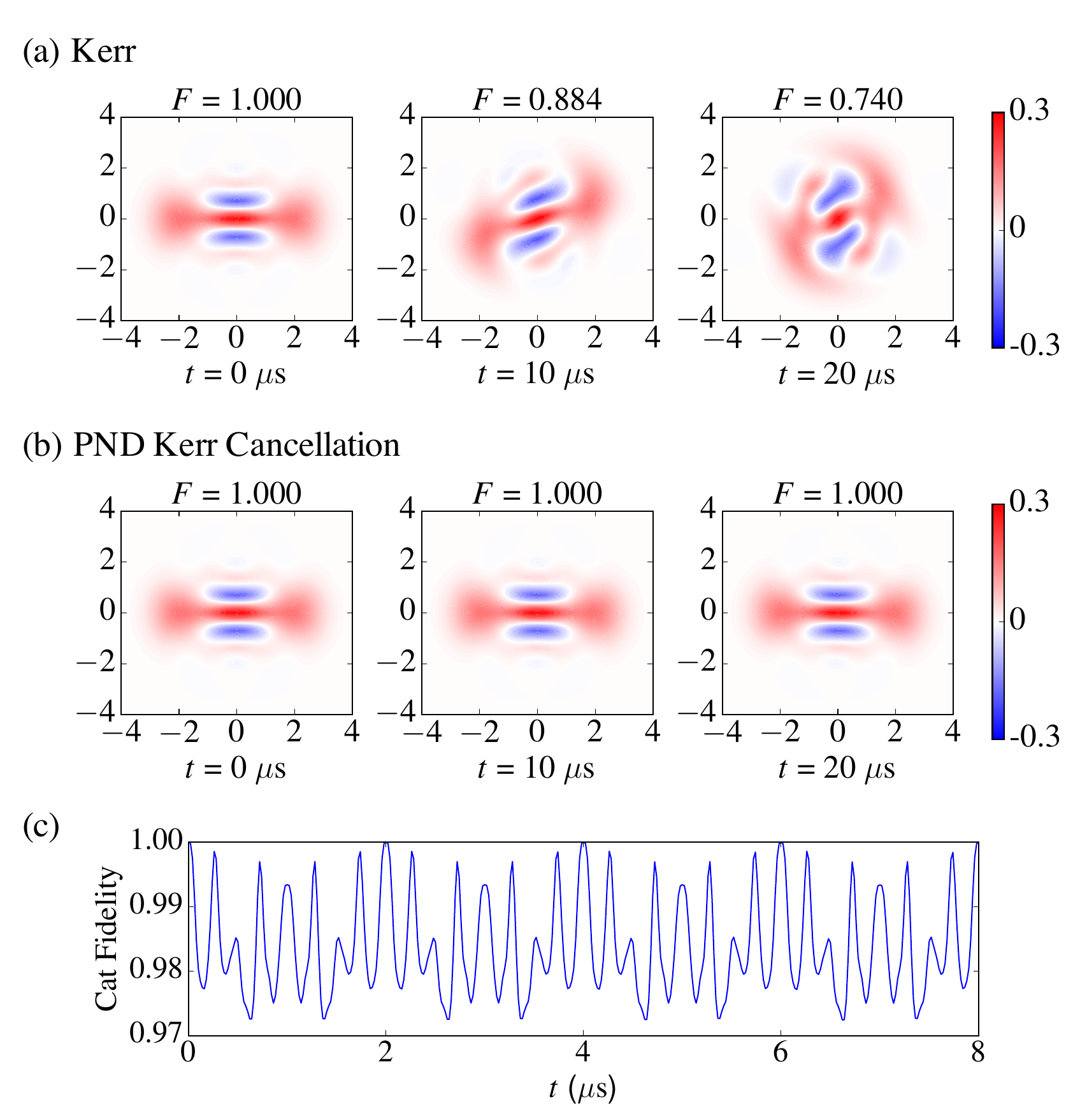}
\caption{Simulated evolution of the cavity cat state $\frac{1}{\sqrt{2}}(\ket{\alpha_c}+\ket{-\alpha_c})$, an even superposition of coherent states with opposite phases. (a) Wigner-function snapshots and the fidelity $F$ of the cat state evolving under the cavity self-Kerr. (b) Wigner-function snapshots and the fidelity $F$ of the cat state with PND Kerr cancellation. (c) Cat-state fidelity as a function of time under PND Kerr cancellation, showing a clear signature of the micromotion.  Here we use parameters in table~\ref{tab:Kerrpara} and assume $\alpha_c=\sqrt{2}$, cavity photon number $N_{\rm cut}=6$, $\chi/2\pi=2$ MHz, and $T_M=2$ $\mu s$.}
\label{fig:Kerr} 
\end{center}
\end{figure}

The effective Hamiltonian seen by the photon while the qubit stays in its ground state is
\begin{align}
H_{\rm  eff,g} =\hbar \omega_a \ad \ah- \frac{\hbar K}{2} \ad\ad\ah\ah+\bra{g} H_{\rm I, eff} \ket{g}.
\label{Heffgkerr}
\end{align}
The self-Kerr effect is the leading-order correction to cavity resonators that can cause unwanted rotations and (in the presence of photon loss can) introduce extra decoherence. We can apply this Kerr-corrected Hamiltonian engineering scheme to cancel the cavity self-Kerr by choosing $\sum_n \hbar E_{T,n} \kb{n}=(\hbar K/2) \ad\ad\ah\ah$, or to engineer a target Hamiltonian while canceling Kerr. Examples of PND parameters with Kerr cancellation are shown in Tables~\ref{tab:Kerrpara} and \ref{tab:ZwithKerr}. Numerical simulation of PND Kerr cancellation is presented in Fig.~\ref{fig:Kerr}. Taking $\chi/2\pi=2$ MHz and $K/2\pi=3$ kHz appropriate for cQED devices~\cite{Heeres2017} and assuming no photon loss, one can preserve a cat state with close to unit fidelity for $t=20\mu$s and 99.2 \% fidelity for $t=100\mu$s with PND Kerr cancellation (see Appendix~\ref{A:Kerr}).

\section{PND Hamiltonian engineering for coupled cavities\label{sec:coupled}}
\begin{figure}[htbp]
\begin{center}
\includegraphics[width=.45 \textwidth]{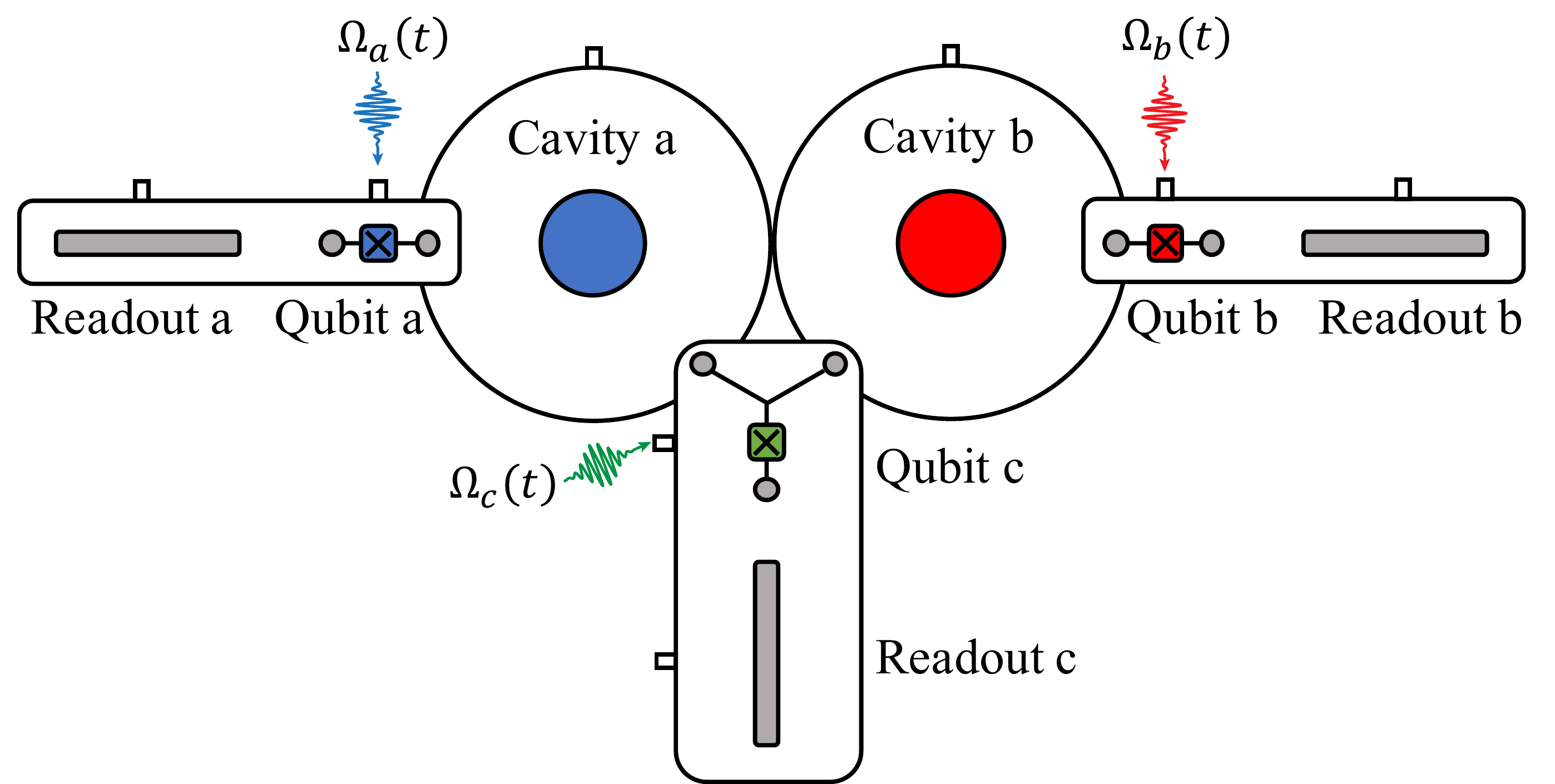}
\caption{Schematic of the cQED implementation of two cavities coupled to three ancilla qubits~\cite{Gao2019}. Two 3D microwave cavity resonators, cavity a and cavity b, are capacitively coupled to two transmon qubits, qubit a and qubit b, respectively.  Another Y-shaped transmon qubit, qubit c, is capavitively coupled to both cavity a and cavity b. Three transmon qubits can be controlled independently through individual drive ports, and they can be coupled to separate readout quasiplanar resonators.}
\label{fig:cqed1} 
\end{center}
\end{figure}

Here we further generalize our PND scheme to the case of coupled cavities.  Specifically, we consider two cavity modes $\ah$ and $\bh$ dispersively coupled to their own ancilla qubits $\hat{\sigma}^a$, $\hat{\sigma}^b$, and to another joint qubit $\hat{\sigma}^c$ with a dispersive shift $\chi_c$ (see Fig.~\ref{fig:cqed1}), assumed to be equal for both modes~\cite{Rosenblum2018},
\begin{align}
\Hh_0&= \hbar \omega_a \ad \ah +\hbar \omega_{q,a} \kb{e_a}- \hbar \chi_a \ad \ah\kb{e_a}\notag\\&+ \hbar \omega_b \bd \bh+\hbar \omega_{q,b} \kb{e_b}- \hbar \chi_b \bd \bh\kb{e_b}\notag\\&+\hbar \omega_{q,c} \kb{e_c}- \hbar \chi_c (\ad \ah+\bd \bh) \kb{e_c},
\label{eqn:Habc}
\end{align}
where $\omega_{a/b}$ are the frequencies of the cavities, $\omega_{q,a/b/c}$ are the qubit transition frequencies between $\ket{g_{a/b/c}}$ and $\ket{e_{a/b/c}}$, and $\chi_{a/b/c}$ are the dispersive coupling strengths.
One can drive the coupled qubit $\hat{\sigma}^c$ to control cavity states dependent on $n_a+n_b$ and drive qubits $\hat{\sigma}^a$ and $\hat{\sigma}^b$ to control cavity states dependent on $n_a$ and $n_b$ respectively. Altogether, one can engineer a two-cavity Hamiltonian $\Hh_{E} =\sum_{n_a, n_b} \hbar E_{n_a n_b} \kb{n_a n_b}=\sum_{n_a, n_b} \hbar (E_{c,n_a+n_b} +E_{a,n_a}+E_{b,n_b}) \kb{n_a n_b}$ (see Appendix~\ref{A:Coupled}).  

We can apply this generalized PND scheme to implement controlled-Z rotations for realizing controlled-phase gates CPHASE$(\theta)$, which is one class of essential two-qubit entangling gates for universal quantum computing.  For bosonic-encoded qubits, the CPHASE gate has been demonstrated fairly recently, though in a protocol susceptible to photon loss during the gate operation~\cite{Xu2020}.  Here we present an error-transparent operation~\cite{Vy2013,Kapit2018} of controlled-Z-rotation by PND which is tolerant against photon loss in the cavities.

\begin{figure}[htbp]
\begin{center}
\includegraphics[width=0.45 \textwidth]{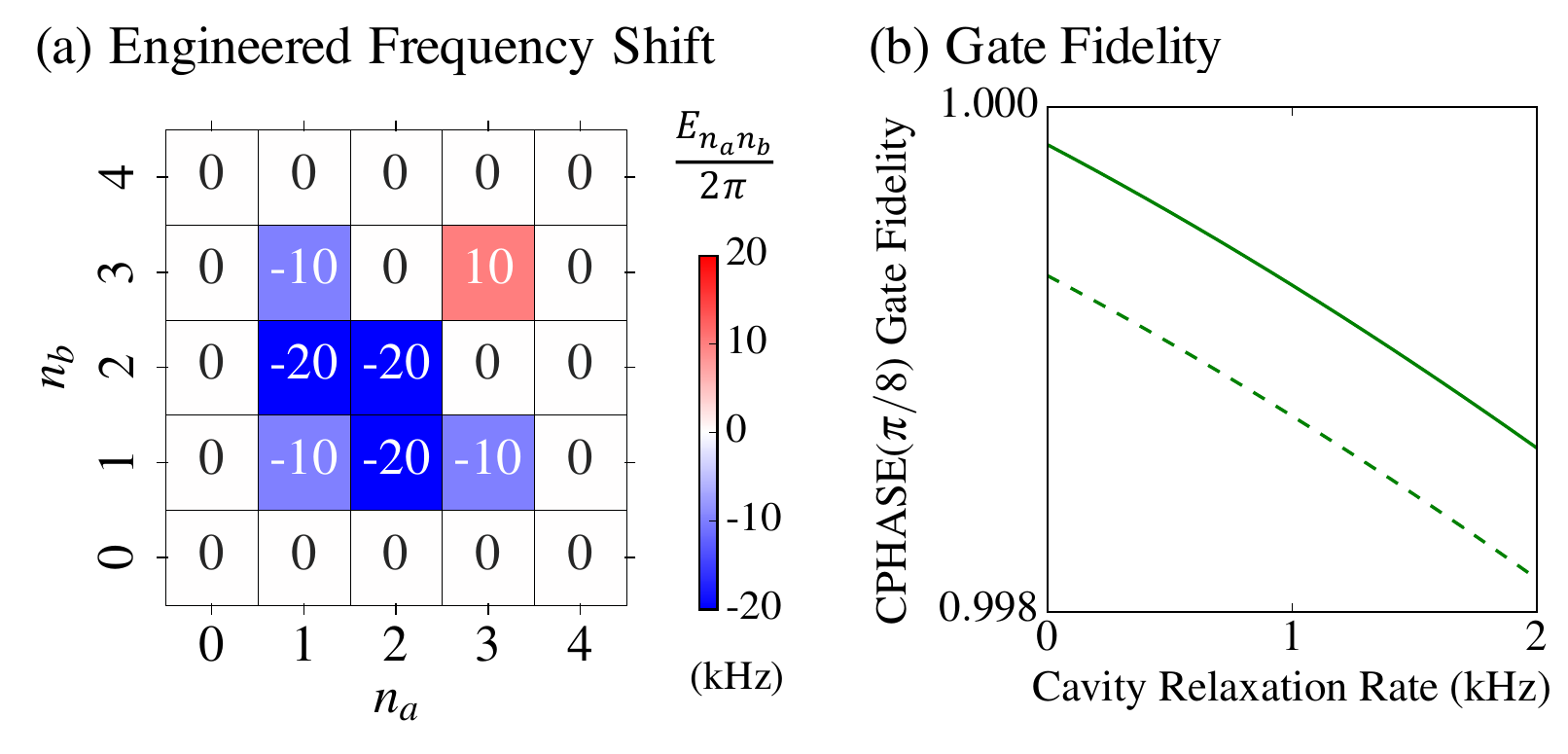}
\caption{(a) Simulated engineered energy shifts for error-transparent controlled-rotation as a function of the photon numbers of coupled cavities. (b) Fidelity of PND CPHASE($\pi/8$) gate as a function of the total cavity relaxation rate.  The dashed line represents the modifications due to ancilla qubit relaxation at a rate $\Gamma_{qa}/2\pi=\Gamma_{qb}/2\pi=\Gamma_{qc}/2\pi=3$ kHz. Here we use parameters in table~\ref{tab:CPHASEa},~\ref{tab:CPHASEb}, and~\ref{tab:CPHASEc} with the CPHASE($\pi/8$) gate time $T_G=16\pi/\chi=2 T_M$, and assume that the two cavities have the same relaxation rate.}
\label{fig:ControlledR} 
\end{center}
\end{figure}

We design an error-transparent Hamiltonian $\Hh_{cR}$ for CPHASE$(\theta)$ such that within a total number distance $d_n =\min(d_{n_a},d_{n_b})$, $\ket{1_a 1_b}_L$ and its error states have the same negative energy shift $\hbar g_{cR},$
\begin{align}
\Hh_{cR} &=-\hbar g_{cR} \sum_{k=0}^{\infty}  \sum_{l_a=0}^{d_{n}-1}\kb{(2 k+1) d_{n_a}-l_a}_a \notag\\
&\otimes  \sum_{l_b=0}^{d_{n}-1-l_a}\kb{(2 k+1) d_{n_b}-l_b}_b,
\end{align}
up to residual energy shifts on error states with total photon loss number exceeding $d_n-1$.
The targeted energy shifts to implement $\Hh_{cR}$ for $d_n=d_{n_a}=d_{n_b}=2$ and the numerically simulated engineered energy shifts by the generalized PND are shown in Fig.~\ref{fig:ControlledR}(a).  The simulated fidelity of a CPHASE($\pi/8$) gate starting from the kitten-code encoded state $(1/2)(\ket{0_a}_k+\ket{1_a}_k) \otimes (\ket{0_b}_k+\ket{1_b}_k)\otimes \ket{g_a g_b g_c}$, followed by instantaneous recovery of single-photon loss in both cavities, is larger than 99.8\% even in the presence of the relaxation of all three ancilla qubits (Fig.~\ref{fig:ControlledR}(b)).

\section{Summary and Outlook\label{sec:summary}}
In conclusion, we develop a toolbox for photon-number-dependent Hamiltonian engineering by off-resonantly driving ancilla qubit(s).   We provide a general formalism to design and optimize the control drives for engineering arbitrary single-cavity target Hamiltonian and performing quantum gates, with examples include three-photon interaction, parity-dependent energy, error-transparent Z rotation for rotation-symmetric bosonic qubits, and cavity self-Kerr cancellation.  We can also generalize this scheme to implement error-transparent controlled rotation between two cavities.  The flexible and thus highly nonlinear engineered Hamiltonian for photons admits versatile applications for quantum simulation and quantum information processing.  Our scheme can be implemented with dispersively coupled microwave cavities and transmon qubits in the cQED platform.  Recent demonstration of the strong dispersive regime in a surface acoustic wave resonator~\cite{Sletten2019,Arrangoiz2019} indicates opportunities for phonon-number-dependent operations as well.

Looking forward, exploring fault-tolerant approaches such as qubit-error-transparent~\cite{Rosenblum2018} or path-independent~\cite{Reinhold2020,Ma2020}  gates may further reduce the decoherence induced by the ancilla qubit.  Robust and continuous control of cavities can assist quantum sensing and realize universal fault-tolerant quantum gates with potential compatibility with autonomous quantum error correction~\cite{Kapit2016,Lihm2018,Sun2020,Lebreuilly2021}.  For future prospects of many-body quantum simulation with photons~\cite{Hartmann2016}, applying our scheme to create local interactions in coupled cavities can offer opportunities for studying exotic phenomena of extended Bose-Hubbard model with three- or more-body interactions~\cite{Dutta2015}.

\begin{acknowledgments}
{We acknowledge support from the ARL-CDQI (W911NF-15-2-0067), ARO (W911NF-18-1-0020, W911NF-18-1-0212), ARO MURI (W911NF-16-1-0349), AFOSR MURI (FA9550-15-1-0015, FA9550-19-1-0399), DOE (DE-SC0019406), NSF (EFMA-1640959, OMA-1936118), and the Packard Foundation (2013-39273).}
\end{acknowledgments}

\appendix
\section{PHYSICAL IMPLEMENTATION OF
THE DISPERSIVE HAMILTONIAN}
In this appendix section we briefly describe a realistic physical implementation of a dispersively coupled qubit-cavity system Hamiltonian, Eq.~(\ref{eqn:H0}), via cQED~\cite{Boissonneault2009}.  Consider a superconducting qubit coupled to a microwave cavity described by a Jaynes-Cummings Hamiltonian
\begin{align}
\Hh_{\rm JC}= \hbar \omega_{a,0} \ad \ah +  \hbar\omega_{q,0} \kb{e} + \hbar g (\ah \shp+ \ad \shm ).
\label{JCCoupling}
\end{align}
Working in the large detuning regime such that $\abs{\Delta}\equiv \abs{\omega_{a,0}-\omega_{q,0}}\gg \abs{g}$, one can apply unitary transformation $\Uh_D=\exp \left[ -\frac{g}{\Delta}(\ah \shp- \ad \shm ) \right]$ to find the perturbative expansion of the Jaynes-Cummings Hamiltonian to the leading-order of $g/\Delta$ as
\begin{align}
\Hh{'}=&\Uh_D\Hh_{\rm JC}\Uh^{\dagger}_D \notag\\=& \hbar \left(\omega_{a,0}+\frac{g^2}{\Delta} \right) \ad \ah +  \hbar \left(\omega_{q,0}-\frac{g^2}{\Delta} \right) \kb{e} \notag\\
&- \hbar \chi \ad \ah \kb{e} +\mathcal{O} (\frac{g^2}{\Delta^2})\notag\\
\approx &\hbar \omega_{a}\ad \ah +  \hbar \omega_{q}\kb{e}  - \hbar \chi \ad \ah \kb{e} = \Hh_0,
\label{JCtoD}
\end{align}
where $\omega_a=\omega_{a,0}+\frac{g^2}{\Delta}$, 
$\omega_q=\omega_{q,0}-\frac{g^2}{\Delta}$, and $\chi = \frac{2g^2}{\Delta}$. We arrive at the dispersive Hamiltonian, which has been extensively explored and utilized in cQED platforms as a key interaction for control and measurement~\cite{Schuster2007,Boissonneault2009,Heeres2015,Heeres2017,Gao2019}.
An example schematic of a transmon superconducting qubit coupled to a 3D microwave cavity is illustrated in Fig.~\ref{fig:cqed}.

\section{EVOLUTION OF THE DRIVEN
DISPERSIVE MODEL \label{A:Dispersive}}
Here we calculate the perturbative expansion of the unitary evolution operator for an off-resonantly driven, dispersively coupled qubit-oscillator system described by the Hamiltonian
\begin{align}
\Hh(t) =& \hbar \omega_a \ad \ah +  \hbar\omega_q \kb{e}- \hbar \chi \ad \ah\kb{e}+ \Vh(t) \notag\\ \equiv  & \Hh_0+ \Vh(t), 
\end{align}
where $\Vh(t)=\hbar \Omega(t) \shm + \hbar \Omega^*(t) \shp$ with $\Omega(t)=\sum_m  \Omega_m e^{i ( \omega_q- m \chi +\delta_m)t}$.   We assume a tri-partition ansatz for the evolution operator $\Uh_{S}(t_f,t_i)$ such that $\ket{\psi(t_f)}\equiv\Uh_{S}(t_f,t_i)\ket{\psi(t_i)}$ for an initial state $\ket{\psi(t_i)}$ and a final state $\ket{\psi(t_f)}$~\cite{Rahav2003} , 
\begin{align}
\Uh_{S}(t_f,t_i)=e^{-i \Gh(t_f)}e^{-i\Hh_{\rm eff}(t_f-t_i)/\hbar}e^{i \Gh(t_i)}, 
\end{align}
where the evolution is separated into a time-independent effective Hamiltonian $\Hh_{\rm eff}$ governing the long-time dynamics, as well as initial and final kicks, $\Gh(t_i)$ and $\Gh(t_f)$.   The subscript $S$ denotes evolution in the Schr\"{o}dinger picture.

Moving to the interaction picture with the unitary transformation $\Uh=\exp(i \Hh_0 t /\hbar)$, we are left with the  term
\begin{align}
\Vh_I(t)=& \Uh \Hh \Uh^{\dagger} - i \hbar \Uh \dot{\Uh}^{\dagger}\notag\\=&\sum_m\sum_n \hbar (\Omega_m e^{i [(n-m) \chi +\delta_m] t} \kb{n} \shm + H.c.).
\end{align}
Here the subscript $I$ denotes operators in the interaction picture, $\hat{O}_I(t)=e^{i \Hh_0 t/\hbar} \hat{O}(t) e^{-i \Hh_0 t/\hbar}$.
The evolution operator in the interaction picture is connected to the schr\"{o}dinger picture one by
\begin{align}
\Uh_I(t_f,t_i)&=e^{i \Hh_0 t_f/\hbar}e^{-i \Gh(t_f)}e^{-i\Hh_{\rm eff}(t_f-t_i)/\hbar}e^{i \Gh(t_i)}e^{-i \Hh_0 t_i/\hbar}\notag\\&=e^{-i\Gh_I(t_f)}e^{-i(\Hh_{\rm eff}-\Hh_0)(t_f-t_i)/\hbar}e^{i\Gh_I(t_i)}.
\end{align}
Assuming $\abs{\Omega_m} \ll \abs{\delta_m} ,\abs{\chi -\delta_m}$, we can use time-dependent perturbation theory to calculate $\Uh_I(t_f,t_i)$ in powers of $\Vh_I$ and find the perturbative expansion of $\Hh_{\rm eff}$ and $\Gh_I(t)$ such that $\Hh_{\rm eff}=\Hhe{0}+\Hhe{1}+\Hh_{\rm eff}^{(2)}+\cdots$ and $\Gh_I(t)=\GhI{0}(t)+\GhI{1}(t)+\GhI{2}(t)+\cdots$.

Specifically,
\begin{align}
\Uh_I(t_f,t_i)=&1-\frac{i}{\hbar} \int_{t_i}^{t_f}\Vh_I(t_1)dt_1\notag\\&+\left(\frac{-i}{\hbar}\right)^2 \int_{t_i}^{t_f}dt_1\int_{t_i}^{t_1}dt_2\Vh_I(t_1)\Vh_I(t_2) +\cdots.
\end{align}
The zeroth order terms are $\GhI{0}(t)=0$ and $\Hhe{0}-\Hh_0=0$. Expanding to $\mathcal{O}(\Vh)$, 
\begin{align}
\int_{t_i}^{t_f}\frac{\Vh_I(t_1)dt_1}{\hbar}= \GhI{1}(t_f)+\frac{\Hhe{1} (t_f-t_i)}{\hbar}- \GhI{1}(t_i).
\end{align}
Since
\begin{align}
\bra{n,g} \int_{t_i}^{t_f}\frac{\Vh_I(t_1)dt_1 }{\hbar}\ket{n,g}=\bra{n,e} \int_{t_i}^{t_f}\frac{\Vh_I(t_1)dt_1}{\hbar} \ket{n,e}=0,
\end{align}
and 
\begin{align}
&\bra{n,g} \int_{t_i}^{t_f}\frac{\Vh_I(t_1)dt_1}{\hbar} \ket{n,e}\notag\\&=\sum_m{\frac{ \Omega_m e^{i((n-m)\chi+\delta_m )t_f}-\Omega_m e^{i((n-m)\chi+\delta_m )t_i}}{i[(n-m)\chi+\delta_m]}},
\end{align}
we have
\begin{align}
\Hhe{1}=0,
\end{align}
\begin{align}
\GhI{1}(t)=\sum_m \sum_n{\frac{ \kb{n}\Omega_m\shm e^{i((n-m)\chi+\delta_m )t}}{i[(n-m)\chi+\delta_m]}}+H.c..
 \label{eqS:G1}
\end{align}

Expanding to $\mathcal{O}(\Vh^2)$, 
\begin{widetext}
\begin{align}
&\left(\frac{-i}{\hbar}\right)^2 \int_{t_i}^{t_f}dt_1\int_{t_i}^{t_1}dt_2\Vh_I(t_1)\Vh_I(t_2)=-i \GhI{2}(t_f) - \frac{i}{\hbar} \Hh_{\rm eff}^{(2)} (t_f-t_i)+i \GhI{2}(t_i)+\frac{1}{2}\left(\frac{-i \Hhe{1} (t_f-t_i)}{\hbar} \right)^2\notag\\&+\frac{1}{\hbar}\Hhe{1} (t_f-t_i) \GhI{1}(t_i)-\frac{1}{\hbar}\GhI{1}(t_f)\Hhe{1} (t_f-t_i)+  \GhI{1}(t_f)\GhI{1}(t_i)- \frac{\GhI{1}(t_i)^2}{2}- \frac{\GhI{1}(t_f)^2}{2}.
\end{align}
We find
\begin{align}
\Hhe{2}=-\sum_m \sum_n \frac{\hbar|\Omega_m|^2 \kb{n}}{(n-m)\chi+\delta_m}\shz,
\end{align}

\begin{align}
\GhI{2}(t)=-\sum_{m_1}\sum_{m_2 \neq m_1} \sum_n \frac{ \Omega_{m_1} \Omega_{m_2}^{*} e^{i(\delta_{m_1}-\delta_{m_2}-(m_1-m_2)\chi)t}-\Omega_{m_1}^{*} \Omega_{m_2} e^{-i(\delta_{m_1}-\delta_{m_2}-(m_1-m_2)\chi)t}}{2 i [(n-m_1)\chi+\delta_{m_1}][\delta_{m_1}-\delta_{m_2}-(m_1-m_2)\chi]}\kb{n}\shz.
\end{align}
The third- and fourth-order terms of the effective Hamiltonian are
\begin{align}
\Hhe{3}=0,
\end{align}
\begin{align}
\Hhe{4}& = \sum_{m_1} \sum_{m_2} \sum_n \frac{\hbar|\Omega_{m_1}|^2 |\Omega_{m_2}|^2 \kb{n} \shz}{[(n-m_1)\chi+\delta_{m_1}][(n-m_2)\chi+\delta_{m_2}]^2}\notag\\
&-\sum_{m_1} \sum_{m_2 \neq m_1}\sum_n\frac{\hbar |\Omega_{m_1}|^2 |\Omega_{m_2}|^2 \kb{n}\shz}{[(n-m_1)\chi+\delta_{m_1}]^2\delta_{m_1}-\delta_{m_2}-(m_1-m_2)\chi]}\notag\\
&+\sum_{m_1,m_2,m_3,m_4} \sum_n\frac{\hbar \Omega_{m_1} \Omega_{m_2}^{*} \Omega_{m_3} \Omega_{m_4}^{*} \kb{n}\shz}{[(n-m_4)\chi+\delta_{m_4}][\delta_{m_1}-\delta_{m_2}-(m_1-m_2)\chi][(n-m_1)\chi+\delta_{m_1}]},
\end{align}
where the last term satisfies the condition  $\delta_{m_1}-m_1\chi+\delta_{m_3}-m_3 \chi=\delta_{m_2}-m_2\chi+\delta_{m_4}-m_4 \chi$, and $m_1 \neq m_2 \neq m_3 \neq m_4$ or $m_1  =m_3 \neq m_2 \neq m_4$ or $m_2  =m_4 \neq m_1 \neq m_3$.
\end{widetext}
Consider special cases that all $\delta_m{'}s$ are commensurate with $\chi$, this problem reduces to a Floquet Hamiltonian with a single 
icity, and one can calculate the Floquet effective Hamiltonian and the kick operator~\cite{Goldman2014} and obtain identical results.

\section{EVOLUTION WITH
QUBIT-INDUCED DEPHASING \label{A:Decoherence}}
Here we consider how errors in the ancilla qubit propagates to the cavity mode under off-resonant drives.  The ancilla errors are described by the qubit relaxation jump operator $\sqrt{\Gamma_q}\shm$ and the qubit dephasing jump operator $\sqrt{\Gamma_{\phi}}\kb{e}$ in the time-dependent Lindblad master equation
\begin{align}
\partial_{t} \rho_{tot} (t)= -\frac{i}{\hbar}[H(t),\rho_{tot}(t)]+ \mathcal{D} (\rho_{tot}(t)),
\end{align}
where $\rho_{tot}$ is the total density matrix of the coupled qubit-oscillator system, and
\begin{align}
&\mathcal{D} (\rho(t))=\sum_k \left[ J_k \rho(t) J_k^{\dagger}-\frac{1}{2}J_k^{\dagger} J_k \rho(t) -\rho(t)\frac{1}{2}J_k^{\dagger} J_k\right],\notag\\
&\{ J_k \}=\{\sqrt{\Gamma_q}\shm, \sqrt{\Gamma_{\phi}}\kb{e},\sqrt{\kappa_a}\ah\}  .
\end{align}
Here $\kappa_a$ represents the relaxation rate of the cavity.

Moving to the interaction picture, $\shm$ becomes $\hat{\sigma}_{-I}(t)=\sum_n \kb{n} e^{-i(\omega_q-n \chi)t}\shm$ while $\kb{e}$ stays the same.  Under the rotating wave approximation, when $\Gamma_q \ll \abs{\chi}$ such that qubit decay will release a photon-number-dependent  energy $\hbar(\omega_q - n \chi)$, we treat the relaxation jump operator as a set of independent jump operators in the cavity number state manifold,
\begin{align}
&\partial_{t} \rho_{tot,I}(t) = -\frac{i}{\hbar}[V_I(t),\rho_{tot,I}(t)]+ \mathcal{D}_{I} (\rho_{tot,I}(t)),
\end{align}
\begin{align}
\mathcal{D}_I (\rho(t))&=\sum_k \left[ J_{k,I} \rho(t) J_{k,I}^{\dagger}-\frac{1}{2}J_{k,I}^{\dagger} J_{k,I} \rho(t) \right. \notag\\& \left. -\rho(t)\frac{1}{2}J_{k,I}^{\dagger} J_{k,I}\right],\notag\\
\{ J_{k,I} \}&=\{\sqrt{\Gamma_{\phi}}\kb{e},\sqrt{\Gamma_q}\shm \kb{n} | \forall n ,\sqrt{\kappa_a}\ah\}.
\end{align}

We now assume again a tri-partition ansatz for the evolution superoperator $\Lambda_{t_f, t_i} $, $\rho_I (t_f) \equiv \Lambda_{t_f, t_i} \rho_I (t_i)$,
\begin{align}
\Lambda_{t_f, t_i}=e^{-\Phi_{t_f}} e^{\tilde{\mathcal{L}}(t_f-t_i)}e^{\Phi_{t_i} },
\end{align}
such that there is a time-independent Liouvillian $\tilde{\mathcal{L}}$ and a kick superoperator $\Phi_t$ that absorbs the time dependence.  For $\Gamma_{\phi}, \Gamma_q, |\Omega_m| \ll \abs{\delta_m},\abs{\chi-\delta_m} $ and $\delta_m \sim \mathcal{O}(\chi)$, one can expand $\tilde{\mathcal{L}}$ and $\Phi_t$ in perturbative orders of $\mathcal{O}(\Omega_m,\Gamma_{\phi}, \Gamma_q)$.

We find the time-independent evolution superoperator as $\tilde{\mathcal{L}} \approx \tilde{\mathcal{L}}^{(1)} + \tilde{\mathcal{L}}^{(2)} + \tilde{\mathcal{L}}^{(3)}$ with
\begin{align}
&\tilde{\mathcal{L}}^{(1)}(\cdot) = \mathcal{D}_I (\cdot), \\
&\tilde{\mathcal{L}}^{(2)}(\cdot)= -\frac{i}{\hbar}\left[\Hh^{(2)}_{\rm eff}, \cdot \right],\\
&\tilde{\mathcal{L}}^{(3)}(\cdot)= \frac{1}{2} \sum_n \sum_m \frac{\left\{ [[\mathcal{S}_{-},\mathcal{D}_I],\mathcal{S}_{+}]+ [[\mathcal{S}_{+},\mathcal{D}_I],\mathcal{S}_{-}]\right\}(\cdot)}{[(n-m)\chi+\delta_m]^2} ,
\end{align}
where $\mathcal{S}_{-}(\cdot)=\left[\kb{n}\Omega_m \shm e^{i [(n-m)\chi +\delta_m]t}, \cdot \right]$, $\mathcal{S}_{+}(\cdot)=\left[\kb{n} \Omega_m^{*} \shp e^{-i [(n-m)\chi +\delta_m]t}, \cdot \right]$,
and the kick superoperator as $\Phi_t \approx \Phi^{(1)}_t + \Phi^{(2)}_t$ with
\begin{align}
&\Phi_t^{(1)}(\cdot)= \left[ i \GhI{1}(t), \cdot \right],\\
&\Phi_t^{(2)}(\cdot)=  \left[ i \GhI{2}(t), \cdot \right] -\sum_n \sum_m \frac{[\mathcal{S}_{+}+\mathcal{S}_{-},\mathcal{D}_I](\cdot)}{[(n-m)\chi+\delta_m]^2} .
\end{align}
Choosing $\delta_m{'}s$ commensurate with $\chi$ such that all the time-dependent terms have a common period $T_M$, then for $t_f=t_i+cT_M$ for some integer $c$, $\rho_I(t_i + c T_M)=e^{-\Phi_{t_i}}e^{\tilde{\mathcal{L}} c T_M} e^{\Phi_{t_i}}=e^{\mathcal{L}_F(t_i) c T_M}$ for an Floquet generator $\mathcal{L}_F(t_i)=e^{-\Phi_{t_i}} \tilde{\mathcal{L}} e^{\Phi_{t_i}}$~\cite{Dai2016,Scopa2019}.  Taking $t_i=0$ and tracing over the ancilla qubit degree of freedom assuming $\rho_{gg}=1$ and $\rho_{ee}=\rho_{ge}=\rho_{eg}=0$, the cavity density matrix in the interaction picture $\rho_{c,I}$ follows a Floquet effective master equation
\begin{align}
\partial_t \rho_{c,I}(t) \approx&  -i \left[ \sum_m \sum_n \frac{|\Omega_m|^2 \kb{n}}{(n-m)\chi+\delta_m}, \rho_{c,I}(t) \right]\notag\\&+ \mathcal{D}_{c,I}( \rho_{c,I}(t)),
\end{align}
with jump operators
\begin{align}
\{ &J_{c,k,I} \}=\left\{ \frac{\sqrt{\Gamma_{\phi}} \Omega_m^{*} \kb{n}}{(n-m)\chi+\delta_m},  \frac{ \sqrt{\Gamma_{q}} \Omega_m^{*} \kb{n} }{(n-m)\chi+\delta_m}| \forall n, m \right\} \notag\\
&\cup \left\{   \sum_n \sum_m \frac{\sqrt{\Gamma_{\phi}} \Omega_m^{*} \kb{n}}{(n-m)\chi+\delta_m}  \right\}  \cup \left\{\sum_m \frac{ \sqrt{\Gamma_{q}} \Omega_m^{*} \kb{n}}{(n-m)\chi+\delta_m} |\forall n \right\}.
\end{align}
The jump operators cause dephasing for off-diagonal density matrix elements of the cavity number states $\rho_{n_1 n_2}$ at a rate
\begin{align} &\gamma_{n_1 n_2}=\frac{\Gamma_{\phi}+\Gamma_{q}}{2} (p_{n_1,e}+p_{n_2,e})
\notag\\&-\Gamma_{\phi} \sum_{m_1}  \frac{\Omega_{m_1}^{*}}{(n_1-m_1)\chi+\delta_{m_1}}\sum_{m_2}  \frac{\Omega_{m_2}}{(n_2-m_2)\chi+\delta_{m_2}},
\end{align}
where $p_{n,e} \equiv \sum_m |\frac{\Omega_m}{(n-m)\chi+\delta_m}|^2+|\sum_m \frac{\Omega_m}{(n-m)\chi+\delta_m}|^2$ is the time-averaged probability of the qubit excited state component $\ket{n,e}$ due to $\GhI{1}(t)$. The second term in $p_{n,e}$, $|\sum_m \frac{\Omega_m}{(n-m)\chi+\delta_m}|^2$, is the contribution from the initial kick.  Smoothly ramping up the drive can thus reduce the qubit-induced dephasing by removing the effect of the kick.

\section{Microscopic Model and Kerr Corrections\label{A:Kerr}}
We now revisit the microscopic model of a resonator mode $\hat{a}$ coupled to another bosonic mode $\hat{d}$ with anharmonicity $\alpha$.  Specifically,
\begin{align}
\Hh=\hbar \omega_{a,0}{\ad \ah} + \hbar \omega_{q,0} \dhd \hd -\frac{\hbar \alpha}{2} \dhd \dhd\hd\hd +\hbar g (\ad \hd + \dhd \ah).
\end{align}
For a small coupling  $g$, one can use perturbation theory to estimate the frequency shifts as a function of photon number in the resonator $n_a$ and the anharmonic mode $n_d$.  Expanding up to the order of $g^4$ and keeping only $n_d=0,1$ (states $\ket{g}$, $\ket{e}$), the generic Hamiltonian of the coupled system reads
\begin{widetext}
\begin{align}
\Hh=&\hbar \left(\omega_{a,0}+\frac{g^2}{\Delta}-\frac{g^4}{\Delta^3}\right){\ad \ah} + \hbar \left(\omega_{q,0} -\frac{g^2}{\Delta}+\frac{g^4}{\Delta^3}\right)\dhd \hd - \frac{2 \hbar g^2 \alpha}{\Delta(\Delta+\alpha)}\ad \ah\dhd\hd-\frac{\hbar g^4 \alpha}{\Delta^3(\alpha+2 \Delta)} \ad\ad\ah\ah \notag\\
&+ \frac{4 \hbar g^4 \alpha (\alpha^2+2\alpha\Delta+2\Delta^2)}{\Delta^3(\Delta+\alpha)^3}\ad \ah\dhd\hd+\frac{\hbar g^4 2 \alpha^2 (3\alpha^3+11\alpha^2 \Delta+15\alpha \Delta^2+9\Delta^3)}{\Delta^3(\alpha+\Delta)^3(\alpha+2\Delta)(3 \alpha+2\Delta)} \ad\ad\ah\ah\dhd\hd +\mathcal{O}(g^6).
\end{align}
Here $\Delta=\omega_a-\omega_q$.

We can reorganize this Hamiltonian in a form by identifying $\dhd \hd$ as $\kb{e}$,
\begin{align}
\Hh =& \hbar \omega_a \ad \ah +  \hbar \omega_q \kb{e} -\hbar \chi\ad \ah \kb{e} - \frac{\hbar K}{2} \ad\ad\ah\ah \notag\\&+\frac{\hbar \chi{'}}{2} \ad\ad\ah\ah\kb{e} ,
\end{align}
where $\omega_a=\omega_{a,0}+\frac{g^2}{\Delta}-\frac{g^4}{\Delta^3}$, $\omega_q=\omega_{q,0}-\frac{g^2}{\Delta}+\frac{g^4}{\Delta^3}$, $\chi=\frac{2 g^2 \alpha}{\Delta(\Delta+\alpha)}-\frac{4 g^4 \alpha (\alpha^2+2\alpha\Delta+2\Delta^2)}{\Delta^3(\Delta+\alpha)^3}$, $K=\frac{2 g^4 \alpha}{\Delta^3(\alpha+2 \Delta)}$, and $\chi{'}=\frac{g^4 4 \alpha^2 (3\alpha^3+11\alpha^2 \Delta+15\alpha \Delta^2+9\Delta^3)}{\Delta^3(\alpha+\Delta)^3(\alpha+2\Delta)(3 \alpha+2\Delta)}$.

Consider an off-resonantly driven coupled system with the photon self-Kerr $K$ and the second-order dispersive shift $\chi{'}$,
\begin{align}
\Hh(t) =& \hbar \omega_a \ad \ah +   \hbar \omega_q \kb{e} -\hbar \chi\ad \ah \kb{e} - \frac{\hbar K}{2} \ad\ad\ah\ah +\frac{\hbar \chi{'}}{2} \ad\ad\ah\ah\kb{e}+ \hbar \Omega(t) \shm+  \hbar \Omega^*(t) \shp \notag\\
 \equiv& \Hh_0 + \Hh_K+ \hbar \Omega(t) \shm+  \hbar \Omega^*(t) \shp, 
\end{align}
here $\Omega(t)=\sum_m  \Omega_m e^{i ( \omega_q- m \chi +\delta_m)t}$.

We again assume a tri-partition ansatz for  the time-evolution operator,
\begin{align}
\Uh_S(t_f,t_i)=e^{-i \Gh(t_f)}e^{-i\Hh_{\rm eff}(t_f-t_i)/\hbar}e^{i \Gh(t_i)}.
\end{align}
Moving to the interaction picture with the unitary transformation $\Uh=\exp(i (\Hh_0+ \Hh_K) t /\hbar)$, we are left with the drive term
\begin{align}
\Vh_I(t)=\sum_m\sum_n \hbar\Omega_m e^{i ((n-m) \chi +\delta_m - \chi{'}n(n-1)/2) t} \kb{n} \shm + H.c..
\end{align}

We can again use the time-dependent perturbation theory to calculate $\Uh_I(t_f,t_i)$ in powers of $\Vh_I$ and find the perturbative expansions $\Hh_{\rm eff}=\Hhe{0}+\Hhe{1}+\Hh_{\rm eff}^{(2)}+\cdots$ and $\Gh_I(t)=\GhI{0}(t)+\GhI{1}(t)+\GhI{2}(t)+\cdots$, with additional contributions from the Kerr term $\Hh_K= \frac{\hbar K}{2} \ad\ad\ah\ah +\frac{\hbar \chi{'}}{2} \ad\ad\ah\ah\kb{e}$.

The zeroth order terms are $\GhI{0}(t)=0$ and $\Hhe{0}-\Hh_0=0$.  We find
\begin{align}
\Hhe{1}=0,
\end{align}
\begin{align}
\GhI{1}(t)=\sum_m \sum_n{\frac{ \Omega_m \kb{n} \shm e^{i[(n-m)\chi -\chi{'}n(n-1)/2+\delta_m]t}}{i[(n-m)\chi- \chi{'}n(n-1)/2+\delta_m]}}+ H.c.,
\end{align}

\begin{align}
\Hhe{2}=-\sum_m \sum_n \frac{\hbar|\Omega_m|^2 \kb{n}}{(n-m)\chi- \chi{'}n(n-1)/2+\delta_m}\shz,
\end{align}
\begin{align}
\GhI{2}=-\sum_{m_1}\sum_{m_2 \neq m_1} \sum_n \frac{\Omega_{m_1} \Omega_{m_2}^{*} \kb{n}\shz  e^{i[\delta_{m_1}-\delta_{m_2}-(m_1-m_2)\chi]t}}{2 i[(n-m_1)\chi- \chi{'}n(n-1)/2+\delta_{m_1}][\delta_{m_1}-\delta_{m_2}-(m_1-m_2)\chi]}+H.c.
\end{align}
The third- and fourth-order terms of the effective Hamiltonian are
\begin{align}
\Hhe{3}=0,
\end{align}
\begin{align}
\Hhe{4}=&\sum_{m_1} \sum_{m_2} \sum_n \frac{\hbar|\Omega_{m_1}|^2 |\Omega_{m_2}|^2 \kb{n} \shz}{[(n-m_1)\chi+\delta_{m_1}- \chi{'}n(n-1)/2][(n-m_2)\chi- \chi{'}n(n-1)+\delta_{m_2}]^2}\notag\\
&-\sum_{m_1}\sum_{m_2 \neq m_1}\sum_n\frac{\hbar |\Omega_{m_1}|^2 |\Omega_{m_2}|^2\kb{n}\shz}{[(n-m_1)\chi- \chi{'}n(n-1)/2+\delta_{m_1}]^2[\delta_{m_1}-\delta_{m_2}-(m_1-m_2)\chi]} \notag\\
&+\frac{\sum_{m_1,m_2,m_3,m_4} \sum_n \hbar \Omega_{m_1} \Omega_{m_2}^{*} \Omega_{m_3} \Omega_{m_4}^{*}\kb{n}\shz}{[(n-m_4)\chi- \chi{'}n(n-1)/2+\delta_{m_4}][\delta_{m_1}-\delta_{m_2}-(m_1-m_2)\chi][(n-m_1)\chi- \chi{'}n(n-1)/2+\delta_{m_1}]},
\end{align}
where the last term satisfies the condition $\delta_{m_1}-m_1\chi+\delta_{m_3}-m_3 \chi=\delta_{m_2}-m_2\chi+\delta_{m_4}-m_4 \chi$, and $m_1 \neq m_2 \neq m_3 \neq m_4$ or $m_1  =m_3 \neq m_2 \neq m_4$ or $m_2  =m_4 \neq m_1 \neq m_3$.

\end{widetext}

\begin{figure}[htbp!]
\begin{center}
\includegraphics[width=0.9 \textwidth]{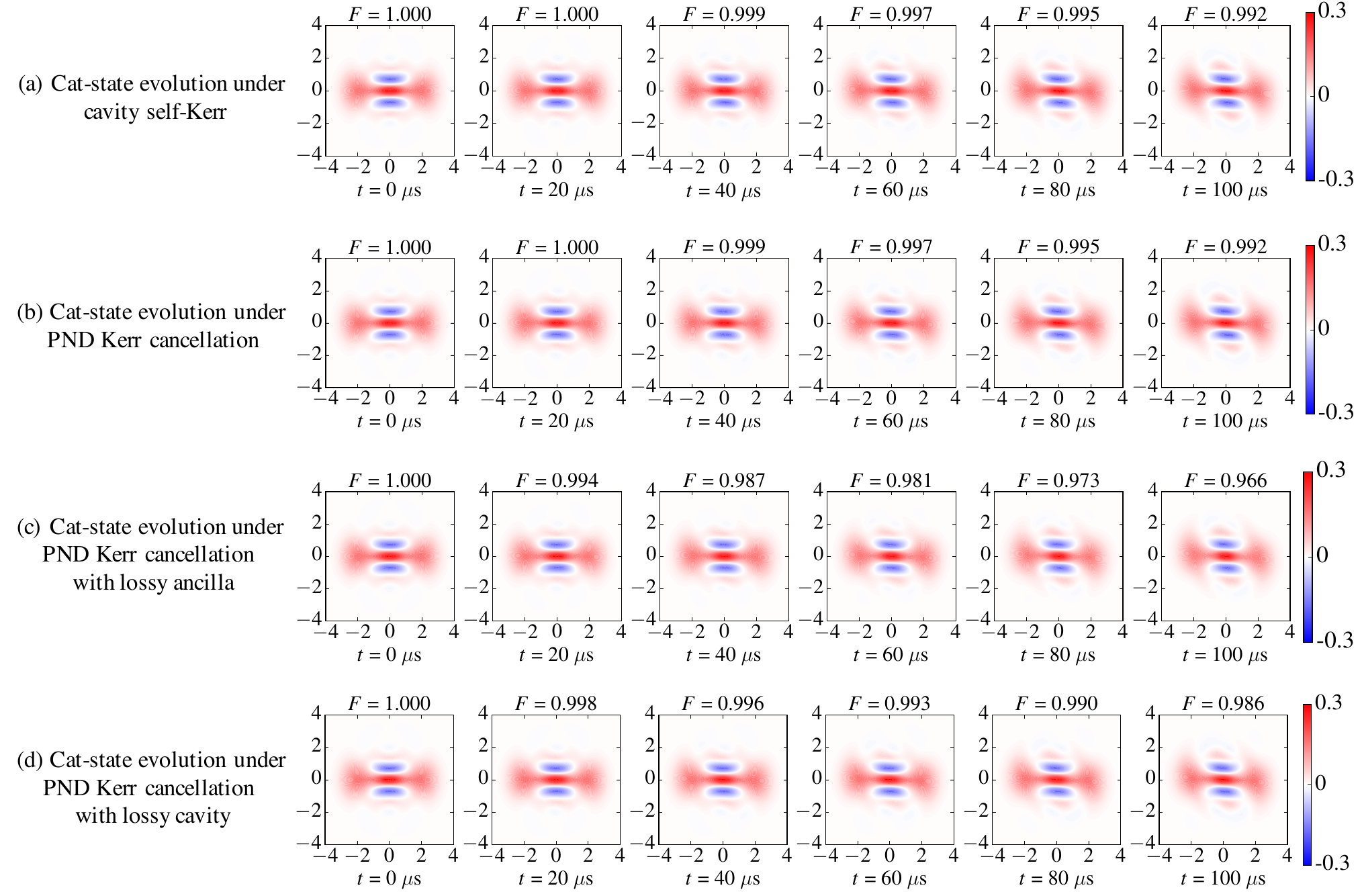}
\caption{Wigner-function snap shots and cat fidelity $F$ under (a) cavity self-Kerr, (b) PND Kerr cancellation, (c) PND Kerr cancellation with a lossy ancilla qubit, and (d) PND Kerr cancellation with a lossy cavity.  
 Here we assume abrupt PND drive with parameters in table~\ref{tab:Kerrpara}, and assume a qubit relaxation rate $\Gamma_q/2\pi=3$ kHz, a cavitiy relaxation rate $\kappa_a/2\pi=0.01$ kHz, and a cat size $\alpha_c=\sqrt{2}$.  The snap shots are chosen at multiples of the micromotion period $T_M=8\pi/\chi=2$ $\mu$s.}
\label{fig:KerrLong} 
\end{center}
\end{figure}

\begin{figure}[htbp]
\begin{center}
\includegraphics[width=0.45 \textwidth]{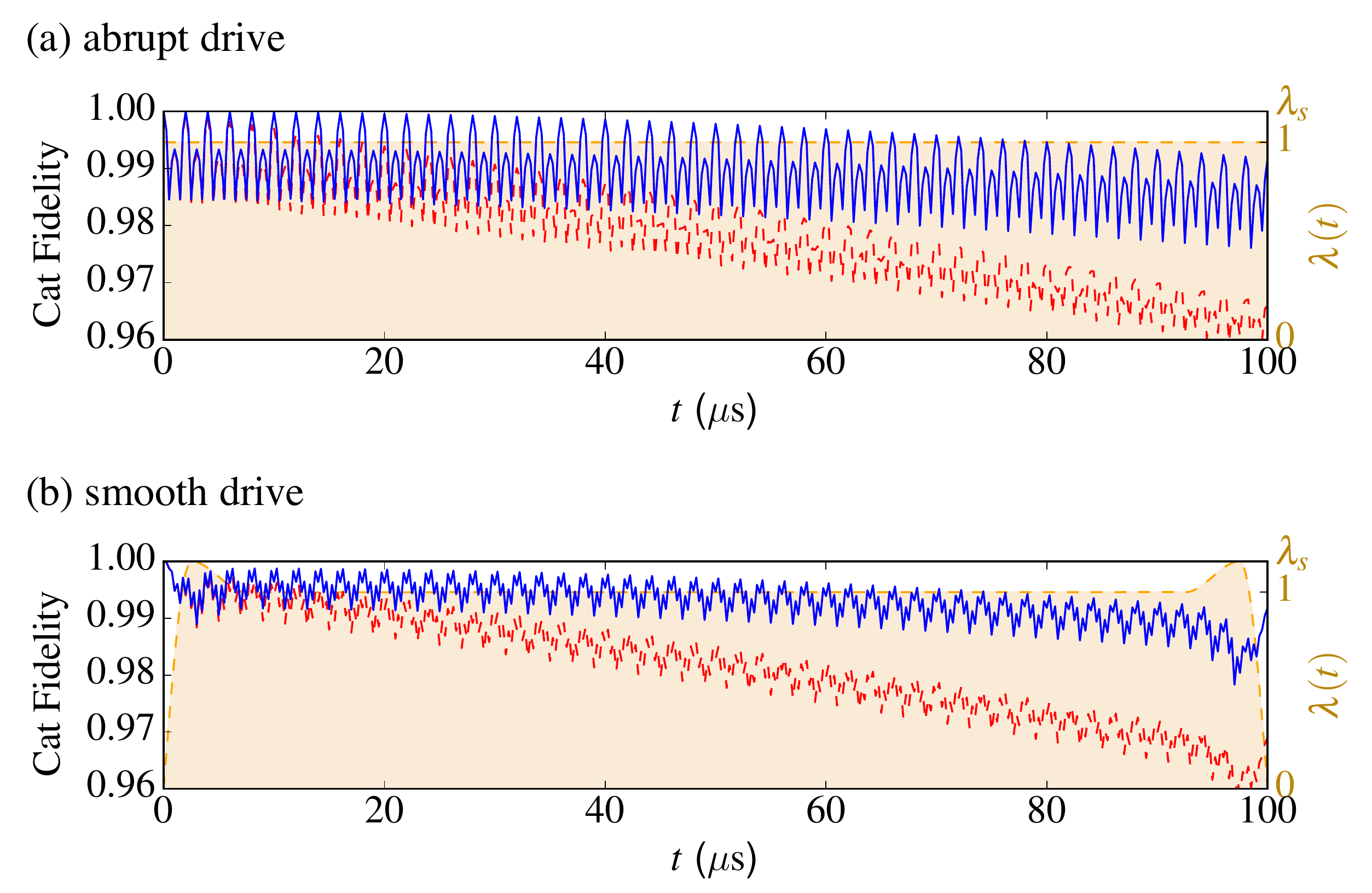}
\caption{Cat-state fidelity during the Kerr cancellation operation by (a) abruptly turning on the PND drive and (b) smoothly turning on and off the PND drive as describe in Eq.~(\ref{eqn:lambdaup}) and ~(\ref{eqn:lambdadown}). The cat fidelity as a function of time is shown in blue (without ancilla qubit relaxation) and red (with ancilla qubit relaxation) curves.  The ramping function $\lambda(t)$ in presented as the filled orange curve. Here we use parameters shown in Table~\ref{tab:Kerrpara}, and assume $\Gamma_q/2\pi=3$ kHz, $\alpha_c=\sqrt{2}$, and $T_s=2.5$ $\mu$s. At the end of the operation, the final cat fidelity is 99.180\% for the abrupt drive and 99.184\% for the smooth drive.  The additional infidelity induced by ancilla relaxation is 2.568\% for the abrupt drive and 2.276\% for the smooth drive.}
\label{fig:smoothKerr} 
\end{center}
\end{figure}

\begin{figure}[htbp]
\begin{center}
\includegraphics[width=0.45 \textwidth]{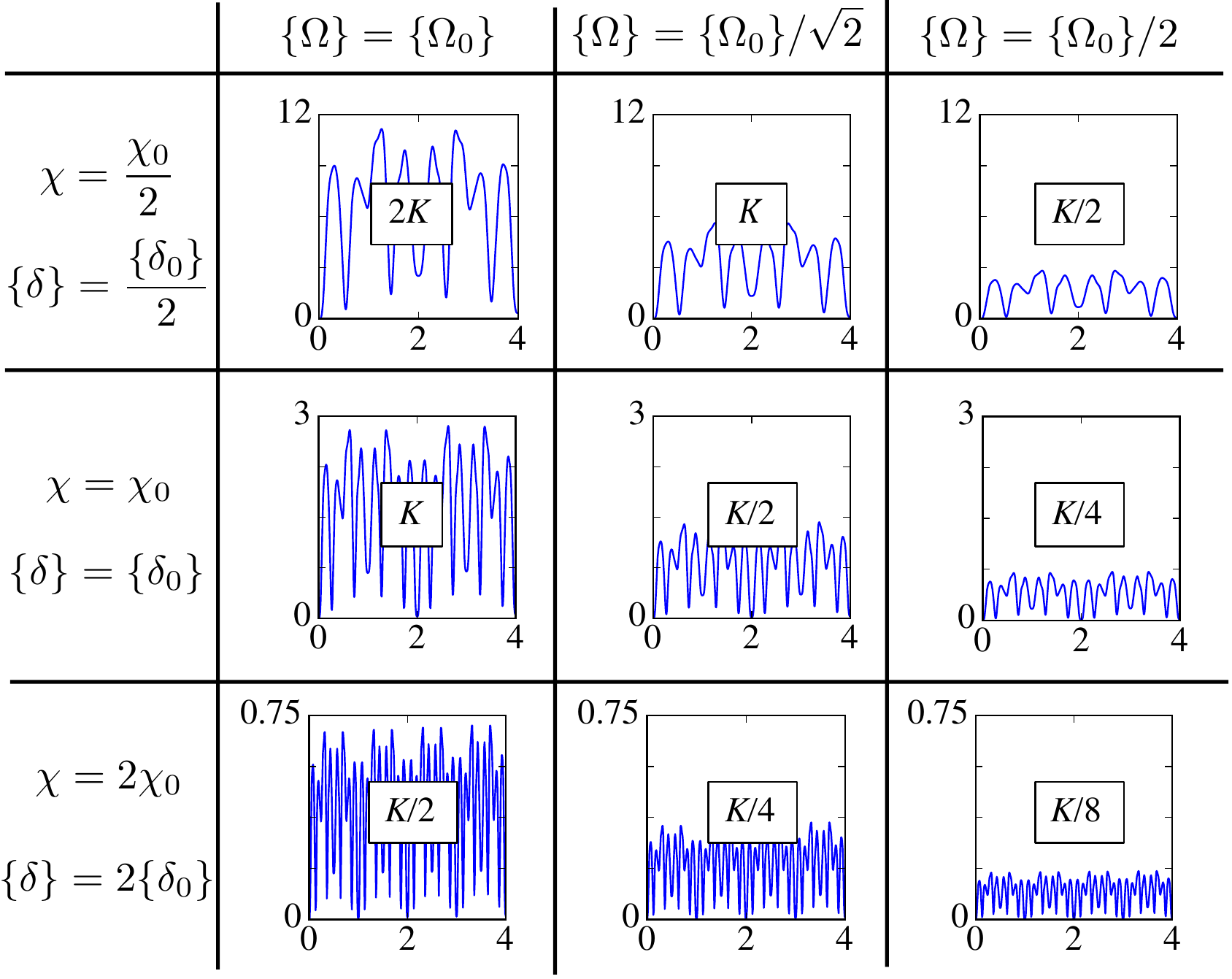}
\caption{Micromotion of the cat-state infidelity under PND Kerr cancellation.  In this table of plots we show the infidelity (y-axis, \%) of a cat state (cat size $\alpha_c=\sqrt{2}$) as a function of time (x-axis, $\mu$s) with varying sets of parameters $\chi, \{ \Omega  \}$, and $\{\delta  \}$. The insets indicate the corresponding cavity self-Kerr value with $K/2\pi=3$ kHz. Here we assume abrupt PND drive, and $\chi_0, \{ \Omega_0  \}$, and  $\{ \delta_0  \}$ are the parameters in Table ~\ref{tab:Kerrpara}. One can clearly infer that the micromotion amplitude scales as $|\Omega^2/\chi^2|$ and that the micromotion period is $T_M=8 \pi/\chi$ as predicted.}
\label{fig:micromotion} 
\end{center}
\end{figure}

\begin{figure}[htbp]
\begin{center}
\includegraphics[width=0.4\textwidth]{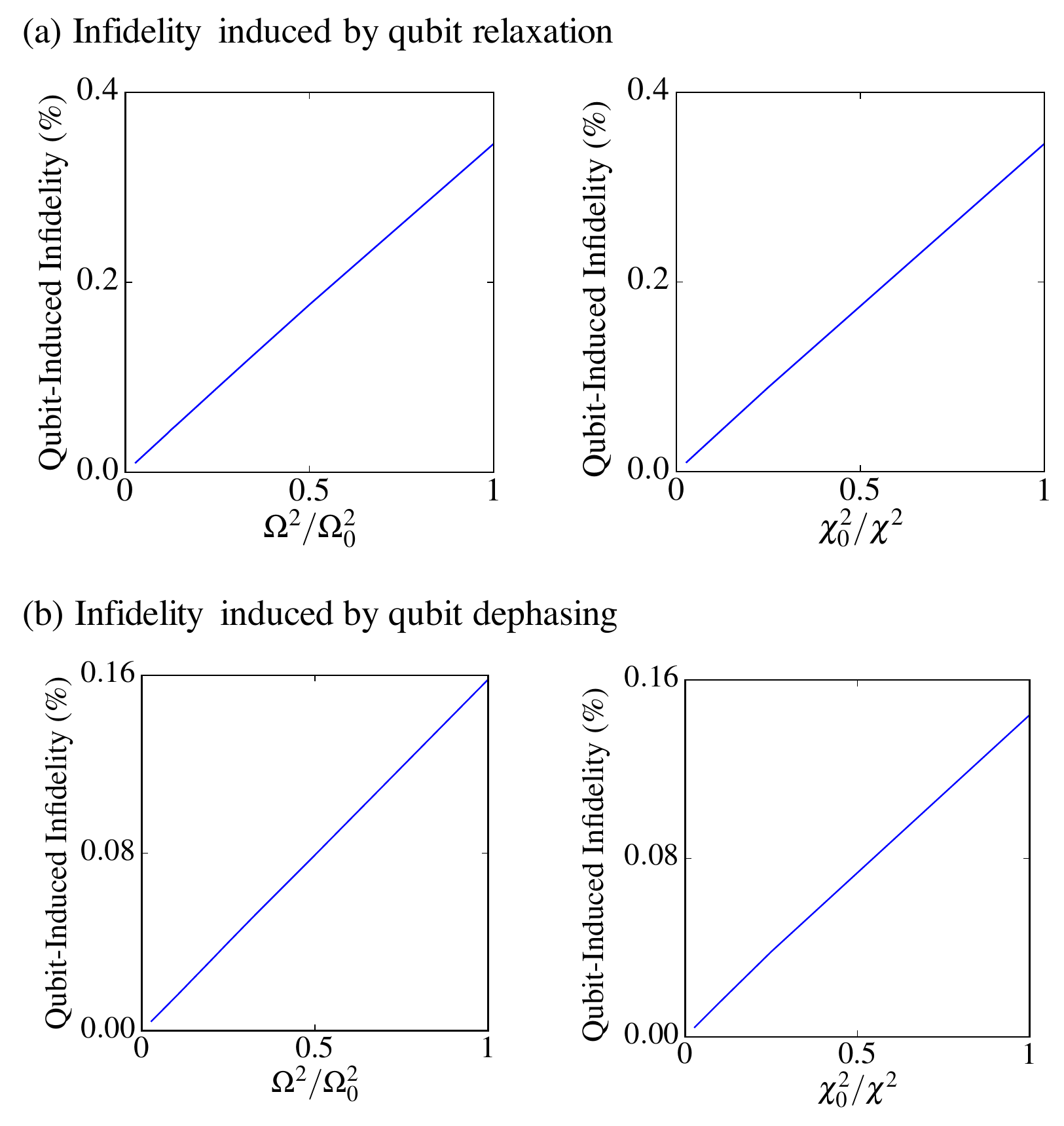}
\caption{Cat infidelity induced by ancilla qubit errors during PND Kerr cancellation.  (a) Cat infidelity at time $t= 12$ $\mu$s induced by qubit relaxation at a rate $\Gamma_q/2 \pi= 3$kHz.  The left panel has a fixed $\Omega=\Omega_0$ with varying $\chi$, and the right panel has a fixed $\chi=\chi_0$ with varying $\Omega$.  (a) Cat infidelity at time $t= 12$ $\mu$s induced by qubit dephasing at a rate $\Gamma_{\phi}/2 \pi= 3 $kHz.  The left panel has a fixed $\Omega=\Omega_0$ with varying $\chi$, and the right panel has a fixed $\chi=\chi_0$ with varying $\Omega$. $\chi_0, \{ \Omega_0  \}$, and $\{ \delta_0  \}$ are the parameters in Table ~\ref{tab:Kerrpara}.  Here we assume abrupt PND drive, $K/2\pi=(\Omega^2 \chi_0/ \Omega_0^2 \chi)3$ kHz, and a cat size $\alpha_c=\sqrt{2}$.}
\label{fig:decoherence}
\end{center}
\end{figure}

Additional simulations of the cat-state evolution under PND Kerr cancellation is shown in Fig. ~\ref{fig:KerrLong} and ~\ref{fig:smoothKerr}. We assume the system starts with an even cavity cat state $(1/\sqrt{2})(\ket{\alpha_c}+\ket{-\alpha_c})$ and the qubit in its ground state then simulate the state evolution in the rotating frame with $\Uh=\exp[i (\omega_a \ad \ah +\omega_q \kb{e}) t]$.  With PND Kerr cancellation and assuming no photon loss, the cat state is preserved at a high fidelity approximately equal to 99.2\% even after a long time $t=100$ $\mu$s.  In Fig. ~\ref{fig:micromotion} and ~\ref{fig:decoherence} we study the micromotion and the qubit-induced infidelity for PND Kerr cancellation.  We find that the amplitude of the micromotion and the qubit-induced infidelity both scale as $\Omega^2/\chi^2$ as predicted.  Note that we use Kerr cancellation with $\chi{'}=2K$ as a special case such that both the cavity self-Kerr and the second-order dispersive shift are cancelled by the engineered Hamiltonian and thus there is a perfect micromotion period $T_M$ unperturbed by the total energy.

\onecolumngrid\
\pagebreak
\twocolumngrid\

\section{HAMILTONIAN ENGINEERING
FOR TWO COUPLED CAVITIES \label{A:Coupled}}
Consider two cavity modes $\ah$ and $\bh$ dispersively coupled to two ancilla qubits $\hat{\sigma}^a$ and $\hat{\sigma}^b$ respectively, and to another qubit $\hat{\sigma}^c$ jointly with a dispersive shift $\chi_c$, assumed to be equal
for both modes,
\begin{align}
\Hh_0&= \hbar \omega_a \ad \ah +\hbar \omega_{q,a} \kb{e_a}- \hbar \chi_a \ad \ah\kb{e_a} \notag\\ &+ \hbar \omega_b \bd \bh+\hbar \omega_{q,b} \kb{e_b}- \hbar \chi_b \bd \bh\kb{e_b}
\notag\\&+\hbar \omega_{q,c} \kb{e_c}- \hbar \chi_c (\ad \ah+\bd \bh) \kb{e_c}.
\label{eqn:Habc}
\end{align}
We can add a drive term 
\begin{align}
\Hh_d(t)&=\hbar \Omega_a(t) \shm^a + \hbar \Omega_a^*(t) \shp^a + \hbar \Omega_b(t) \shm^b \notag\\&+ \hbar \Omega_b^*(t) \shp^b + \hbar \Omega_c(t) \shm^c + \hbar \Omega_c^*(t) \shp^c
\label{eqn:Hdabc}
\end{align}
with $\Omega_a(t)=\sum_{m_a}  \Omega_{m_a} e^{i ( \omega_{q,a}- m_a \chi_a +\delta_{m_a})t}$, $\Omega_b(t)=\sum_{m_b}  \Omega_{m_b} e^{i ( \omega_{q,b}- m_b \chi_b +\delta_{m_b})t}$, and $\Omega_c(t)=\sum_{m_c}  \Omega_{m_c} e^{i ( \omega_{q,c}- m_c \chi_c +\delta_{m_c})t}$ to engineer Hamiltonian for the two coupled cavities, assuming $|\Omega_{m_i}| \ll \chi_i, |\delta_{m_i}|$, $i=a,b,c$.
Specifically, the effective Hamiltonian will be of the form $\Hh_{\rm eff}=\Hhe{a}+\Hhe{b}+\Hhe{c}$ with second order terms
\begin{align}
\Hhe{a,2}=-\sum_{m_a} \sum_{n_a} \frac{\hbar|\Omega_{m_a}|^2 \kb{n_a}}{(n_a-m_a)\chi_a+\delta_{m_a}}\shz^a,
\end{align}
\begin{align}
\Hhe{b,2}=-\sum_{m_b} \sum_{n_b} \frac{\hbar|\Omega_{m_b}|^2 \kb{n_b}}{(n_b-m_b)\chi_b+\delta_{m_b}}\shz^b,
\end{align}
\begin{align}
\Hhe{c,2}=-\sum_{m_c} \sum_{n_a} \sum_{n_b} \frac{\hbar|\Omega_{m_c}|^2 \kb{n_a,n_b}}{(n_a+n_b-m_c)\chi_c+\delta_{m_c}}\shz^c.
\end{align}
The fourth order terms can be found as in section I.  The engineered Hamiltonian is 
\begin{align}
\Hh_{\rm E} &=\bra{g_a g_b g_c} \Hh_{\rm eff} \ket{g_a g_b g_c} =\sum_{n_a, n_b} \hbar E_{n_a n_b} \kb{n_a n_b} \notag\\&=\sum_{n_a, n_b} \hbar (E_{c,n_a+n_b} +E_{a,n_a}+E_{b,n_b}) \kb{n_a n_b}\notag\\
& \approx \sum_{m_a} \sum_{n_a} \frac{\hbar|\Omega_{m_a}|^2 \kb{n_a}}{(n_a-m_a)\chi_a+\delta_{m_a}}\notag\\&+\sum_{m_b} \sum_{n_b} \frac{\hbar|\Omega_{m_b}|^2  \kb{n_b}}{(n_b-m_b)\chi_b+\delta_{m_b}}\notag\\&+\sum_{m_c} \sum_{n_a} \sum_{n_b} \frac{\hbar|\Omega_{m_c}|^2 \kb{n_a,n_b}}{(n_a+n_b-m_c)\chi_c+\delta_{m_c}}.
\end{align}
\subsection{Error-transparent controlled-Z rotation}
The form of the engineered Hamiltonian does not have the full degree of freedom to create arbitrary structure of $E_{n_a n_b}$ but is enough to design error-transparent controlled rotation along the Z axis for implementing CPHASE gate on rotational-symmetric bosonic codes.  Specifically, by solving for target frequency shifts
\begin{align}
E_{T,n_a}= \begin{cases}
\, \,\, \, g_{cR}/4 \quad n_a\mod (2d_{n_a})=& 0,2d_{n_a}-1, \notag\\& \cdots, d_{n_a}+1\\
-g_{cR}/4 \quad n_a\mod (2d_{n_a})=& d_{n_a},d_{n_a}-1, \notag\\ & \cdots, 1
\end{cases},
\end{align}
which takes
\begin{align}
&\{I, \hat{a}, \cdots \hat{a}^{d_{n_a}-1}\}\ket{0_a}_L \rightarrow e^{-i g_{cR}t/4} \{I, \hat{a}, \cdots \hat{a}^{d_{n_a}-1}\}\ket{0_a}_L, \notag\\ &\{I, \hat{a}, \cdots \hat{a}^{d_{n_a}-1}\}\ket{1_a}_L \rightarrow e^{i g_{cR}t/4} \{I, \hat{a}, \cdots \hat{a}^{d_{n_a}-1}\}\ket{1_a}_L,
\end{align}
\begin{align}
E_{T,n_b}= \begin{cases}
\, \,\, \, g_{cR}/4 \quad n_b\mod (2d_{n_b})=& 0,2d_{n_b}-1, \notag\\& \cdots, d_{n_b}+1\\
-g_{cR}/4 \quad n_a\mod (2d_{n_b})=& d_{n_b},d_{n_b}-1, \notag\\ & \cdots, 1
\end{cases},
\end{align}
which takes
\begin{align}
&\{I, \hat{b}, \cdots \hat{b}^{d_{n_b}-1}\}\ket{0_b}_L \rightarrow e^{-i g_{cR}t/4} \{I, \hat{b}, \cdots \hat{b}^{d_{n_b}-1}\}\ket{0_b}_L, \notag\\ &\{I, \hat{b}, \cdots \hat{b}^{d_{n_b}-1}\}\ket{1_b}_L \rightarrow e^{i g_{cR}t/4} \{I, \hat{b}, \cdots \hat{b}^{d_{n_b}-1}\}\ket{1_b}_L,
\end{align}
and
\begin{align}
E_{T,n_a+n_b}= \begin{cases}
-g_{cR}/2 & n_a+n_b\mod (d_{n_a}+d_{n_b})\notag\\&=0,d_{n_a}+d_{n_b}-1, \cdots \notag\\& \, \, , d_{n_a}+d_{n_b}-d_n+1\\
\quad \quad 0 & \text{otherwise}
\end{cases},
\end{align}
which takes
\begin{align}
&\{I, \hat{a}^{l_a}\hat{b}^{l_b}\}\ket{0_a 0_b}_L \rightarrow e^{i g_{cR}t/2}\{I, \hat{a}^{l_a}\hat{b}^{l_b}\}\ket{0_a 0_b}_L,
\notag\\&\{I, \hat{a}^{l_a}\hat{b}^{l_b}\}\ket{0_a 1_b}_L \rightarrow\{I, \hat{a}^{l_a}\hat{b}^{l_b}\} \ket{0_a 1_b}_L,\notag\\& \{I, \hat{a}^{l_a}\hat{b}^{l_b}\}\ket{1_a 0_b}_L \rightarrow \{I, \hat{a}^{l_a}\hat{b}^{l_b}\}\ket{1_a 0_b}_L ,\notag\\& \{I, \hat{a}^{l_a}\hat{b}^{l_b}\}\ket{1_a 1_b}_L \rightarrow e^{i g_{cR}t/2} \{I, \hat{a}^{l_a}\hat{b}^{l_b}\}\ket{1_a 1_b}_L,
\end{align}
for $l_a=0, 1, \cdots, d_n-1$ and $l_b=0, 1, \cdots, d_n-1-l_a$. Within a total number distance $d_n$=min($d_{n_a},d_{n_b}$), the overall Hamiltonian should accumulate the same phase for $\ket{1_a 1_b}_L$ and its error states while keeping $\ket{0_a 0_b}_L$, $\ket{0_a 1_b}_L$, $\ket{1_a 0_b}_L$ and their error states unchanged.

\section{OPTIMIZED PND PARAMETERS \label{A:parameters}}
In this section we show tables of optimized PND Hamiltonian engineering parameters that minimizes $\sum_n p_{n,e}$.  All the engineered frequency shifts are subject to a Fourier transformation precision of $\pm 0.5$ kHz. Here we show optimized parameters with real $\Omega_m${'}s. We can also relax the condition and solve for complex values of $\Omega_m${'}s, which give rise to similar performance.

\begin{widetext}

\begin{table}[htbp!]
\caption{Three-Photon Interaction: $\chi/2\pi=2.56$ MHz, $K_3/2\pi=0.5$ kHz (Fig.~\ref{fig:simulations}(a) red)}
\begin{ruledtabular}
\begin{tabular}{ c c c c c c c c }
photon number  & n=0 &n=1&n=2&n=3&n=4&n=5&n=6\\
 \hline
$E_{T,n}/2\pi$ (kHz)   & 0    & 0 &   0&3 & 12& 30& 60\\
$E_n/2\pi$ (kHz) & 0    & 0 &   0&3 & 12& 30& 60\\
$\delta_n/\chi$& 1/2 & 1/2 & 1/2 & 1/2 & 1/2 & 1/4 & 1/2\\
$\Omega_n/\chi$ &0.0946 & 0.0694 &0.0637 &0.0640 &0.0661 &0.0704 &0.0859\\
\end{tabular}
\end{ruledtabular}
\label{tab:threephoton1} 
\end{table}

\begin{table}[htbp!]
\caption{Three-Photon Interaction: $\chi/2\pi=2.56$ MHz, $K_3/2\pi=1$ kHz (Fig.~\ref{fig:simulations}(a) blue)}
\begin{ruledtabular}
\begin{tabular}{ c c c c c c c c }
photon number  & n=0 &n=1&n=2&n=3&n=4&n=5&n=6\\
\hline
target $E_{T,n}/2\pi$ (kHz)   & 0    & 0 &   0&6 & 24& 60& 120\\
engineered $E_n/2\pi$ (kHz) & 0    & 0 &  -1 & 8 & 25 & 61& 122\\
$\delta_n/\chi$& 1/2 & 1/2 & 1/2 & 1/2 & 1/4 & 1/2 & 1/2\\
$\Omega_n/\chi$ & 0.1422 &0.1025&0.0935& 0.0917&0.0995&0.1337&0.1172\\
\end{tabular}
\end{ruledtabular}
\label{tab:threephoton2} 
\end{table}

\begin{table}[htbp!]
\caption{Parity-Dependent Energy: $\chi/2\pi=2.56$ MHz, $P=20$ kHz (Fig.~\ref{fig:simulations}(b) red)}
\begin{ruledtabular}
\begin{tabular}{ c c c c c c c c }
photon number  & n=0 &n=1&n=2&n=3&n=4&n=5&n=6\\
   \hline
target $E_{T,n}/2\pi$ (kHz)   & -20    & 20 &   -20&20 & -20& 20 & -20\\
engineered $E_n/2\pi$ (kHz) & -20  &  20  & -20 &  20 & -20 & 20 & -20 \\
$\delta_n/\chi$& -1/4 & 1/4 & -1/2 & 1/4 & -1/4 & 1/4 & -1/2\\
$\Omega_n/\chi$ & 0.00682 & 0.0568 & 0.0553 & 0.0349 & 0.0427 & 0.0427 & 0.0786\\
\end{tabular}
\end{ruledtabular}
\label{tab:parity1} 
\end{table}

\begin{table}[htbp!]
\caption{Parity-Dependent Energy: $\chi/2\pi=2.56$ MHz, $P=40$ kHz (Fig.~\ref{fig:simulations}(b) blue)}
\begin{ruledtabular}
\begin{tabular}{ c c c c c c c c }
photon number  & n=0 &n=1&n=2&n=3&n=4&n=5&n=6\\
   \hline
target $E_{T,n}/2\pi$ (kHz)   & -40    & 40 &   -40 & 40 & -40& 40 & -40\\
engineered $E_n/2\pi$ (kHz) & -40.5 & 40.5 & -40.5&  40.5& -40.5&  40.5& -40.5\\
$\delta_n/\chi$& -1/2 & 1/4 & -1/2 & 1/4 & -1/4 & 1/2 & -1/4\\
$\Omega_n/\chi$ &0.0232 & 0.0799 & 0.0826 & 0.0463& 0.0469 & 0.0820 & 0.0816\\
\end{tabular}
\end{ruledtabular}
\label{tab:parity2} 
\end{table}

\begin{table}[htbp!]
\caption{Error-Transparent Z-rotation: $\chi/2\pi=2.56$ MHz, $g_R/2\pi=20$ kHz, $d_n$=2 (Fig.~\ref{fig:simulations}(c) red, Fig.~\ref{fig:simulations}(d))}
\begin{ruledtabular}
\begin{tabular}{ c c c c c c c c }
 photon number  & n=0 &n=1&n=2&n=3&n=4&n=5&n=6\\
 \hline
target $E_{T,n}/2\pi$ (kHz)   & 20& -20& -20&  20&  20&-20& -20\\
engineered $E_n/2\pi$ (kHz) & 20& -20& -20&  20&  20&-20& -20\\
$\delta_n/\chi$& 1/2 & -1/2 & -1/2 & 1/4 & 1/2 & -1/4 & -1/2\\
$\Omega_n/\chi$ &0.0862 & 0.0531& 0.0753  &0.0240& 0.0554& 0.0489& 0.0893\\
\end{tabular}
\end{ruledtabular}
\label{tab:piovereight} 
\end{table}

\begin{table}[htbp!]
\caption{Error-Transparent Z-rotation: $\chi/2\pi=2.56$ MHz, $g_R/2\pi=40$ kHz, $d_n$=2 (Fig.~\ref{fig:simulations}(c) blue)}
\begin{ruledtabular}
\begin{tabular}{ c c c c c c c c }
photon number  & n=0 &n=1&n=2&n=3&n=4&n=5&n=6\\
\hline
 target $E_{T,n}/2\pi$ (kHz)   & 40 & -40 & -40 &  40 &  40 & -40 & -40\\
engineered $E_n/2\pi$ (kHz) &  40 & -41 & -41 &  40 &  40 & -41 & -40\\
$\delta_n/\chi$& 1/2 & -1/4 & -1/2 & 1/4 & 1/4 & -1/4 & -1/2\\
$\Omega_n/\chi$ &0.1166&0.0600&-0.0961&0.0308&0.0629&0.0678&0.1214
\end{tabular}
\end{ruledtabular}
\label{tab:piovereight2} 
\end{table} 

\begin{table}[htbp!]
\caption{Kerr Cancellation: $\chi/2\pi=2$ MHz, $K/2\pi=3$ kHz, ${\chi}'/2\pi=6$ kHz (Fig.~\ref{fig:Kerr}(b) and Fig.~\ref{fig:Kerr}(c))}
\begin{ruledtabular}
\begin{tabular}{ c c c c c c c c }
photon number  & n=0 &n=1&n=2&n=3&n=4&n=5&n=6\\
 \hline
$E_{T,n}/2\pi$ (kHz)   & 0    & 0 & 3  &9 & 18& 30& 45\\
$E_n/2\pi$ (kHz) & 0    & 0 &   3 &9 & 18& 30.25 & 46.25\\
$\delta_n/\chi$& 1/2 & 1/2 & 1/2 & 1/2 & 1/2 & 1/4 & 1/4\\
$\Omega_n/\chi$ & 0.0883 & 0.0658 & 0.0635 &  0.0639&  0.0620 & 0.0534 & 0.0606\\
\end{tabular}
\end{ruledtabular}
\label{tab:Kerrpara} 
\end{table} 

\begin{table}[htbp!]
\caption{Error-transparent Z-rotation with Kerr Cancellation: $\chi/2\pi=2$ MHz, $g_R= 20$ kHz, $K/2\pi=3$ kHz, ${\chi}'/2\pi=6$ kHz}
\begin{ruledtabular}
\begin{tabular}{ c c c c c c c c }
photon number  & n=0 &n=1&n=2&n=3&n=4&n=5&n=6\\
 \hline
$E_{T,n}/2\pi$ (kHz)   & 20 & -20 & -17 & 29 & 38 & 10 & 25\\
$E_n/2\pi$ (kHz) & 20 & -20 & -17 & 29 & 38 & 9 & 24\\
$\delta_n/\chi$& 1/2 &- 1/2 & -1/4 & 1/2 & 1/4 & 1/2 & 1/2\\
$\Omega_n/\chi$ & 0.0949 & 0.0659 & 0.0344 & 0.0838 & 0.0588 & 0.0257 & 0.0527\\
\end{tabular}
\end{ruledtabular}
\label{tab:ZwithKerr} 
\end{table} 

\begin{table}[htbp!]
\caption{Error-transparent controlled-Z rotation (drives on qubit $\hat{\sigma}^a$): $\chi_a/2\pi=2.56$ MHz}
\begin{ruledtabular}
\begin{tabular}{ c c c c c c }
photon number  & $n_a$=0 &$n_a$=1&$n_a$=2&$n_a$=3&$n_a$=4\\
 \hline
$E_{T,n_a}/2\pi$ (kHz)   & 5 &-5& -5& 5& 5\\
$E_{a,n_a}/2\pi$ (kHz) &  5 &-5& -5& 5& 5\\
$\delta_{n_a}/\chi$& 1/2 & -1/4 & -1/2 &-1/2&-1/2\\
$\Omega_{n_a}/\chi$ & 0.0393& 0.0212& 0.0365&0.0243& 0.0175\\
\end{tabular}
\end{ruledtabular}
\label{tab:CPHASEa}
\end{table} 

\begin{table}[htbp!]
\caption{Error-transparent controlled-Z rotation (drives on qubit $\hat{\sigma}^b$): $\chi_b/2\pi=2.56$ MHz}
\begin{ruledtabular}
\begin{tabular}{ c c c c c c }
photon number  $n_b$ & $n_b$=0 &$n_b$=1&$n_b$=2&$n_b$=3&$n_b$=4\\
 \hline
$E_{T,n_b}/2\pi$ (kHz)   & 5 &-5& -5& 5& 5\\
$E_{b,n_b}/2\pi$ (kHz) &  5 &-5& -5& 5& 5\\
$\delta_{n_b}/\chi$& 1/2 & -1/4 & -1/2 &-1/2&-1/2\\
$\Omega_{n_b}/\chi$ & 0.0393& 0.0212& 0.0365&0.0243& 0.0175\\
\end{tabular}
\end{ruledtabular}
\label{tab:CPHASEb}
\end{table} 

\begin{table}[htbp!]
\caption{Error-transparent controlled-Z rotation (drives on qubit $\hat{\sigma}^c$): $\chi_c/2\pi=2.56$ MHz}
\begin{ruledtabular}
\begin{tabular}{ c c c c c c c c c c}
photon number $n_c=n_a+n_b$  & $n_c$=0 &$n_c$=1&$n_c$=2&$n_c$=3&$n_c$=4&$n_c$=5&$n_c$=6&$n_c$=7&$n_c$=8\\
 \hline
$E_{T,n_b}/2\pi$ (kHz)   &  -10 & 0 & 0& -10& -10& 0 & 0& -10& -10\\
$E_{b,n_b}/2\pi$ (kHz) &  -10 & 0 & 0& -10& -10& 0 & 0& -10& -10\\
$\delta_{n_b}/\chi$& -1/2& 1/4& 1/2& -1/4& -1/2& -1/4& -1/4& -1/2& -1/2\\
$\Omega_{n_b}/\chi$ & 0.0280 & 0.0197 & 0.0268& 0.0245 & 0.0421 &0.0257 &0.00486 &0.0379 &0.0633\\
\end{tabular}
\end{ruledtabular}
\label{tab:CPHASEc}
\end{table} 

\end{widetext}

\bibliography{PNDRef}

\begin{thebibliography}{46}%
\makeatletter
\providecommand \@ifxundefined [1]{%
 \@ifx{#1\undefined}
}%
\providecommand \@ifnum [1]{%
 \ifnum #1\expandafter \@firstoftwo
 \else \expandafter \@secondoftwo
 \fi
}%
\providecommand \@ifx [1]{%
 \ifx #1\expandafter \@firstoftwo
 \else \expandafter \@secondoftwo
 \fi
}%
\providecommand \natexlab [1]{#1}%
\providecommand \enquote  [1]{``#1''}%
\providecommand \bibnamefont  [1]{#1}%
\providecommand \bibfnamefont [1]{#1}%
\providecommand \citenamefont [1]{#1}%
\providecommand \href@noop [0]{\@secondoftwo}%
\providecommand \href [0]{\begingroup \@sanitize@url \@href}%
\providecommand \@href[1]{\@@startlink{#1}\@@href}%
\providecommand \@@href[1]{\endgroup#1\@@endlink}%
\providecommand \@sanitize@url [0]{\catcode `\\12\catcode `\$12\catcode
  `\&12\catcode `\#12\catcode `\^12\catcode `\_12\catcode `\%12\relax}%
\providecommand \@@startlink[1]{}%
\providecommand \@@endlink[0]{}%
\providecommand \url  [0]{\begingroup\@sanitize@url \@url }%
\providecommand \@url [1]{\endgroup\@href {#1}{\urlprefix }}%
\providecommand \urlprefix  [0]{URL }%
\providecommand \Eprint [0]{\href }%
\providecommand \doibase [0]{http://dx.doi.org/}%
\providecommand \selectlanguage [0]{\@gobble}%
\providecommand \bibinfo  [0]{\@secondoftwo}%
\providecommand \bibfield  [0]{\@secondoftwo}%
\providecommand \translation [1]{[#1]}%
\providecommand \BibitemOpen [0]{}%
\providecommand \bibitemStop [0]{}%
\providecommand \bibitemNoStop [0]{.\EOS\space}%
\providecommand \EOS [0]{\spacefactor3000\relax}%
\providecommand \BibitemShut  [1]{\csname bibitem#1\endcsname}%
\let\auto@bib@innerbib\@empty
\bibitem [{\citenamefont {Schoelkopf}\ and\ \citenamefont
  {Girvin}(2008)}]{Schoelkopf2008}%
  \BibitemOpen
  \bibfield  {author} {\bibinfo {author} {\bibfnamefont {R.~J.}\ \bibnamefont
  {Schoelkopf}}\ and\ \bibinfo {author} {\bibfnamefont {S.~M.}\ \bibnamefont
  {Girvin}},\ }\bibfield  {title} {\enquote {\bibinfo {title} {{Wiring up
  quantum systems}},}\ }\href {\doibase 10.1038/451664a} {\bibfield  {journal}
  {\bibinfo  {journal} {Nature (London)}\ }\textbf {\bibinfo {volume} {451}},\
  \bibinfo {pages} {664} (\bibinfo {year} {2008})}\BibitemShut {NoStop}%
\bibitem [{\citenamefont {Devoret}\ and\ \citenamefont
  {Schoelkopf}(2013)}]{Devoret2013}%
  \BibitemOpen
  \bibfield  {author} {\bibinfo {author} {\bibfnamefont {M.~H.}\ \bibnamefont
  {Devoret}}\ and\ \bibinfo {author} {\bibfnamefont {R.~J.}\ \bibnamefont
  {Schoelkopf}},\ }\bibfield  {title} {\enquote {\bibinfo {title}
  {{Superconducting circuits for quantum information: An outlook}},}\ }\href
  {\doibase 10.1126/science.1231930} {\bibfield  {journal} {\bibinfo  {journal}
  {Science}\ }\textbf {\bibinfo {volume} {339}},\ \bibinfo {pages} {1169}
  (\bibinfo {year} {2013})}\BibitemShut {NoStop}%
\bibitem [{\citenamefont {Kjaergaard}\ \emph {et~al.}(2019)\citenamefont
  {Kjaergaard}, \citenamefont {Schwartz}, \citenamefont {Braum{\"{u}}ller},
  \citenamefont {Krantz}, \citenamefont {Wang}, \citenamefont {Gustavsson},\
  and\ \citenamefont {Oliver}}]{Kjaergaard2019}%
  \BibitemOpen
  \bibfield  {author} {\bibinfo {author} {\bibfnamefont {Morten}\ \bibnamefont
  {Kjaergaard}}, \bibinfo {author} {\bibfnamefont {Mollie~E.}\ \bibnamefont
  {Schwartz}}, \bibinfo {author} {\bibfnamefont {Jochen}\ \bibnamefont
  {Braum{\"{u}}ller}}, \bibinfo {author} {\bibfnamefont {Philip}\ \bibnamefont
  {Krantz}}, \bibinfo {author} {\bibfnamefont {Joel I-Jan}\ \bibnamefont
  {Wang}}, \bibinfo {author} {\bibfnamefont {Simon}\ \bibnamefont
  {Gustavsson}}, \ and\ \bibinfo {author} {\bibfnamefont {William~D.}\
  \bibnamefont {Oliver}},\ }\bibfield  {title} {\enquote {\bibinfo {title}
  {{Superconducting Qubits: Current State of Play}},}\ }\href
  {http://arxiv.org/abs/1905.13641} {\bibfield  {journal} {\bibinfo  {journal}
  {Annu. Rev. Condens. Matter Phys.}\ }\textbf {\bibinfo {volume} {11}},\
  \bibinfo {pages} {369} (\bibinfo {year} {2019})}\BibitemShut {NoStop}%
\bibitem [{\citenamefont {Lei}\ \emph {et~al.}(2020)\citenamefont {Lei},
  \citenamefont {Krayzman}, \citenamefont {Ganjam}, \citenamefont {Frunzio},\
  and\ \citenamefont {Schoelkopf}}]{Lei2020}%
  \BibitemOpen
  \bibfield  {author} {\bibinfo {author} {\bibfnamefont {Chan~U.}\ \bibnamefont
  {Lei}}, \bibinfo {author} {\bibfnamefont {Lev}\ \bibnamefont {Krayzman}},
  \bibinfo {author} {\bibfnamefont {Suhas}\ \bibnamefont {Ganjam}}, \bibinfo
  {author} {\bibfnamefont {Luigi}\ \bibnamefont {Frunzio}}, \ and\ \bibinfo
  {author} {\bibfnamefont {Robert~J.}\ \bibnamefont {Schoelkopf}},\ }\bibfield
  {title} {\enquote {\bibinfo {title} {{High coherence superconducting
  microwave cavities with indium bump bonding}},}\ }\href {\doibase
  10.1063/5.0003907} {\bibfield  {journal} {\bibinfo  {journal} {Appl. Phys.
  Lett.}\ }\textbf {\bibinfo {volume} {116}},\ \bibinfo {pages} {154002}
  (\bibinfo {year} {2020})}\BibitemShut {NoStop}%
\bibitem [{\citenamefont {Gottesman}\ \emph {et~al.}(2001)\citenamefont
  {Gottesman}, \citenamefont {Kitaev},\ and\ \citenamefont
  {Preskill}}]{Gottesman2001}%
  \BibitemOpen
  \bibfield  {author} {\bibinfo {author} {\bibfnamefont {Daniel}\ \bibnamefont
  {Gottesman}}, \bibinfo {author} {\bibfnamefont {Alexei}\ \bibnamefont
  {Kitaev}}, \ and\ \bibinfo {author} {\bibfnamefont {John}\ \bibnamefont
  {Preskill}},\ }\bibfield  {title} {\enquote {\bibinfo {title} {{Encoding a
  qubit in an oscillator}},}\ }\href {\doibase 10.1103/PhysRevA.64.012310}
  {\bibfield  {journal} {\bibinfo  {journal} {Phys. Rev. A}\ }\textbf {\bibinfo
  {volume} {64}},\ \bibinfo {pages} {012310} (\bibinfo {year}
  {2001})}\BibitemShut {NoStop}%
\bibitem [{\citenamefont {Mirrahimi}\ \emph {et~al.}(2014)\citenamefont
  {Mirrahimi}, \citenamefont {Leghtas}, \citenamefont {Albert}, \citenamefont
  {Touzard}, \citenamefont {Schoelkopf}, \citenamefont {Jiang},\ and\
  \citenamefont {Devoret}}]{Mirrahimi2014}%
  \BibitemOpen
  \bibfield  {author} {\bibinfo {author} {\bibfnamefont {Mazyar}\ \bibnamefont
  {Mirrahimi}}, \bibinfo {author} {\bibfnamefont {Zaki}\ \bibnamefont
  {Leghtas}}, \bibinfo {author} {\bibfnamefont {Victor~V.}\ \bibnamefont
  {Albert}}, \bibinfo {author} {\bibfnamefont {Steven}\ \bibnamefont
  {Touzard}}, \bibinfo {author} {\bibfnamefont {Robert~J.}\ \bibnamefont
  {Schoelkopf}}, \bibinfo {author} {\bibfnamefont {Liang}\ \bibnamefont
  {Jiang}}, \ and\ \bibinfo {author} {\bibfnamefont {Michel~H.}\ \bibnamefont
  {Devoret}},\ }\bibfield  {title} {\enquote {\bibinfo {title} {{Dynamically
  protected cat-qubits: A new paradigm for universal quantum computation}},}\
  }\href {\doibase 10.1088/1367-2630/16/4/045014} {\bibfield  {journal}
  {\bibinfo  {journal} {New J. Phys.}\ }\textbf {\bibinfo {volume} {16}},\
  \bibinfo {pages} {045014} (\bibinfo {year} {2014})}\BibitemShut {NoStop}%
\bibitem [{\citenamefont {Michael}\ \emph {et~al.}(2016)\citenamefont
  {Michael}, \citenamefont {Silveri}, \citenamefont {Brierley}, \citenamefont
  {Albert}, \citenamefont {Salmilehto}, \citenamefont {Jiang},\ and\
  \citenamefont {Girvin}}]{Michael2016}%
  \BibitemOpen
  \bibfield  {author} {\bibinfo {author} {\bibfnamefont {Marios~H}\
  \bibnamefont {Michael}}, \bibinfo {author} {\bibfnamefont {Matti}\
  \bibnamefont {Silveri}}, \bibinfo {author} {\bibfnamefont {R~T}\ \bibnamefont
  {Brierley}}, \bibinfo {author} {\bibfnamefont {Victor~V}\ \bibnamefont
  {Albert}}, \bibinfo {author} {\bibfnamefont {Juha}\ \bibnamefont
  {Salmilehto}}, \bibinfo {author} {\bibfnamefont {Liang}\ \bibnamefont
  {Jiang}}, \ and\ \bibinfo {author} {\bibfnamefont {S~M}\ \bibnamefont
  {Girvin}},\ }\bibfield  {title} {\enquote {\bibinfo {title} {{New class of
  quantum error-correcting codes for a bosonic mode}},}\ }\href {\doibase
  10.1103/PhysRevX.6.031006} {\bibfield  {journal} {\bibinfo  {journal} {Phys.
  Rev. X}\ }\textbf {\bibinfo {volume} {6}},\ \bibinfo {pages} {031006}
  (\bibinfo {year} {2016})}\BibitemShut {NoStop}%
\bibitem [{\citenamefont {Albert}\ \emph {et~al.}(2018)\citenamefont {Albert},
  \citenamefont {Noh}, \citenamefont {Duivenvoorden}, \citenamefont {Young},
  \citenamefont {Brierley}, \citenamefont {Reinhold}, \citenamefont {Vuillot},
  \citenamefont {Li}, \citenamefont {Shen}, \citenamefont {Girvin},
  \citenamefont {Terhal},\ and\ \citenamefont {Jiang}}]{Albert2018}%
  \BibitemOpen
  \bibfield  {author} {\bibinfo {author} {\bibfnamefont {Victor~V.}\
  \bibnamefont {Albert}}, \bibinfo {author} {\bibfnamefont {Kyungjoo}\
  \bibnamefont {Noh}}, \bibinfo {author} {\bibfnamefont {Kasper}\ \bibnamefont
  {Duivenvoorden}}, \bibinfo {author} {\bibfnamefont {Dylan~J.}\ \bibnamefont
  {Young}}, \bibinfo {author} {\bibfnamefont {R.~T.}\ \bibnamefont {Brierley}},
  \bibinfo {author} {\bibfnamefont {Philip}\ \bibnamefont {Reinhold}}, \bibinfo
  {author} {\bibfnamefont {Christophe}\ \bibnamefont {Vuillot}}, \bibinfo
  {author} {\bibfnamefont {Linshu}\ \bibnamefont {Li}}, \bibinfo {author}
  {\bibfnamefont {Chao}\ \bibnamefont {Shen}}, \bibinfo {author} {\bibfnamefont
  {S.~M.}\ \bibnamefont {Girvin}}, \bibinfo {author} {\bibfnamefont
  {Barbara~M.}\ \bibnamefont {Terhal}}, \ and\ \bibinfo {author} {\bibfnamefont
  {Liang}\ \bibnamefont {Jiang}},\ }\bibfield  {title} {\enquote {\bibinfo
  {title} {{Performance and structure of single-mode bosonic codes}},}\ }\href
  {\doibase 10.1103/PhysRevA.97.032346} {\bibfield  {journal} {\bibinfo
  {journal} {Phys. Rev. A}\ }\textbf {\bibinfo {volume} {97}},\ \bibinfo
  {pages} {032346} (\bibinfo {year} {2018})}\BibitemShut {NoStop}%
\bibitem [{\citenamefont {Grimsmo}\ \emph {et~al.}(2020)\citenamefont
  {Grimsmo}, \citenamefont {Combes},\ and\ \citenamefont
  {Baragiola}}]{Grimsmo2019}%
  \BibitemOpen
  \bibfield  {author} {\bibinfo {author} {\bibfnamefont {Arne~L.}\ \bibnamefont
  {Grimsmo}}, \bibinfo {author} {\bibfnamefont {Joshua}\ \bibnamefont
  {Combes}}, \ and\ \bibinfo {author} {\bibfnamefont {Ben~Q.}\ \bibnamefont
  {Baragiola}},\ }\bibfield  {title} {\enquote {\bibinfo {title} {{Quantum
  Computing with Rotation-Symmetric Bosonic Codes}},}\ }\href {\doibase
  10.1103/PhysRevX.10.011058} {\bibfield  {journal} {\bibinfo  {journal} {Phys.
  Rev. X}\ }\textbf {\bibinfo {volume} {10}},\ \bibinfo {pages} {011058}
  (\bibinfo {year} {2020})}\BibitemShut {NoStop}%
\bibitem [{\citenamefont {Grimm}\ \emph {et~al.}(2020)\citenamefont {Grimm},
  \citenamefont {Frattini}, \citenamefont {Puri}, \citenamefont {Mundhada},
  \citenamefont {Touzard}, \citenamefont {Mirrahimi}, \citenamefont {Girvin},
  \citenamefont {Shankar},\ and\ \citenamefont {Devoret}}]{Grimm2020}%
  \BibitemOpen
  \bibfield  {author} {\bibinfo {author} {\bibfnamefont {A}~\bibnamefont
  {Grimm}}, \bibinfo {author} {\bibfnamefont {N~E}\ \bibnamefont {Frattini}},
  \bibinfo {author} {\bibfnamefont {S}~\bibnamefont {Puri}}, \bibinfo {author}
  {\bibfnamefont {S~O}\ \bibnamefont {Mundhada}}, \bibinfo {author}
  {\bibfnamefont {S}~\bibnamefont {Touzard}}, \bibinfo {author} {\bibfnamefont
  {M}~\bibnamefont {Mirrahimi}}, \bibinfo {author} {\bibfnamefont {S~M}\
  \bibnamefont {Girvin}}, \bibinfo {author} {\bibfnamefont {S}~\bibnamefont
  {Shankar}}, \ and\ \bibinfo {author} {\bibfnamefont {M~H}\ \bibnamefont
  {Devoret}},\ }\bibfield  {title} {\enquote {\bibinfo {title} {{Stabilization
  and operation of a Kerr-cat qubit}},}\ }\href {\doibase
  10.1038/s41586-020-2587-z} {\bibfield  {journal} {\bibinfo  {journal} {Nature
  (London)}\ }\textbf {\bibinfo {volume} {584}},\ \bibinfo {pages} {205}
  (\bibinfo {year} {2020})}\BibitemShut {NoStop}%
\bibitem [{\citenamefont {Ofek}\ \emph {et~al.}(2016)\citenamefont {Ofek},
  \citenamefont {Petrenko}, \citenamefont {Heeres}, \citenamefont {Reinhold},
  \citenamefont {Leghtas}, \citenamefont {Vlastakis}, \citenamefont {Liu},
  \citenamefont {Frunzio}, \citenamefont {Girvin}, \citenamefont {Jiang},
  \citenamefont {Mirrahimi}, \citenamefont {Devoret},\ and\ \citenamefont
  {Schoelkopf}}]{Ofek2016}%
  \BibitemOpen
  \bibfield  {author} {\bibinfo {author} {\bibfnamefont {Nissim}\ \bibnamefont
  {Ofek}}, \bibinfo {author} {\bibfnamefont {Andrei}\ \bibnamefont {Petrenko}},
  \bibinfo {author} {\bibfnamefont {Reinier}\ \bibnamefont {Heeres}}, \bibinfo
  {author} {\bibfnamefont {Philip}\ \bibnamefont {Reinhold}}, \bibinfo {author}
  {\bibfnamefont {Zaki}\ \bibnamefont {Leghtas}}, \bibinfo {author}
  {\bibfnamefont {Brian}\ \bibnamefont {Vlastakis}}, \bibinfo {author}
  {\bibfnamefont {Yehan}\ \bibnamefont {Liu}}, \bibinfo {author} {\bibfnamefont
  {Luigi}\ \bibnamefont {Frunzio}}, \bibinfo {author} {\bibfnamefont {S.~M.}\
  \bibnamefont {Girvin}}, \bibinfo {author} {\bibfnamefont {L.}~\bibnamefont
  {Jiang}}, \bibinfo {author} {\bibfnamefont {Mazyar}\ \bibnamefont
  {Mirrahimi}}, \bibinfo {author} {\bibfnamefont {M.~H.}\ \bibnamefont
  {Devoret}}, \ and\ \bibinfo {author} {\bibfnamefont {R.~J.}\ \bibnamefont
  {Schoelkopf}},\ }\bibfield  {title} {\enquote {\bibinfo {title} {{Extending
  the lifetime of a quantum bit with error correction in superconducting
  circuits}},}\ }\href {\doibase 10.1038/nature18949} {\bibfield  {journal}
  {\bibinfo  {journal} {Nature (London)}\ }\textbf {\bibinfo {volume} {536}},\
  \bibinfo {pages} {441} (\bibinfo {year} {2016})}\BibitemShut {NoStop}%
\bibitem [{\citenamefont {Hartmann}(2016)}]{Hartmann2016}%
  \BibitemOpen
  \bibfield  {author} {\bibinfo {author} {\bibfnamefont {Michael~J.}\
  \bibnamefont {Hartmann}},\ }\bibfield  {title} {\enquote {\bibinfo {title}
  {{Quantum simulation with interacting photons}},}\ }\href {\doibase
  10.1088/2040-8978/18/10/104005} {\bibfield  {journal} {\bibinfo  {journal}
  {J. Opt.}\ }\textbf {\bibinfo {volume} {18}},\ \bibinfo {pages} {104005}
  (\bibinfo {year} {2016})}\BibitemShut {NoStop}%
\bibitem [{\citenamefont {Noh}\ and\ \citenamefont
  {Angelakis}(2017)}]{Noh2017}%
  \BibitemOpen
  \bibfield  {author} {\bibinfo {author} {\bibfnamefont {Changsuk}\
  \bibnamefont {Noh}}\ and\ \bibinfo {author} {\bibfnamefont {Dimitris~G.}\
  \bibnamefont {Angelakis}},\ }\bibfield  {title} {\enquote {\bibinfo {title}
  {{Quantum simulations and many-body physics with light}},}\ }\href {\doibase
  10.1088/0034-4885/80/1/016401} {\bibfield  {journal} {\bibinfo  {journal}
  {Reports Prog. Phys.}\ }\textbf {\bibinfo {volume} {80}},\ \bibinfo {pages}
  {016401} (\bibinfo {year} {2017})}\BibitemShut {NoStop}%
\bibitem [{\citenamefont {Wang}\ \emph {et~al.}(2020)\citenamefont {Wang},
  \citenamefont {Curtis}, \citenamefont {Lester}, \citenamefont {Zhang},
  \citenamefont {Gao}, \citenamefont {Freeze}, \citenamefont {Batista},
  \citenamefont {Vaccaro}, \citenamefont {Chuang}, \citenamefont {Frunzio},
  \citenamefont {Jiang}, \citenamefont {Girvin},\ and\ \citenamefont
  {Schoelkopf}}]{Wang2020}%
  \BibitemOpen
  \bibfield  {author} {\bibinfo {author} {\bibfnamefont {Christopher~S.}\
  \bibnamefont {Wang}}, \bibinfo {author} {\bibfnamefont {Jacob~C.}\
  \bibnamefont {Curtis}}, \bibinfo {author} {\bibfnamefont {Brian~J.}\
  \bibnamefont {Lester}}, \bibinfo {author} {\bibfnamefont {Yaxing}\
  \bibnamefont {Zhang}}, \bibinfo {author} {\bibfnamefont {Yvonne~Y.}\
  \bibnamefont {Gao}}, \bibinfo {author} {\bibfnamefont {Jessica}\ \bibnamefont
  {Freeze}}, \bibinfo {author} {\bibfnamefont {Victor~S.}\ \bibnamefont
  {Batista}}, \bibinfo {author} {\bibfnamefont {Patrick~H.}\ \bibnamefont
  {Vaccaro}}, \bibinfo {author} {\bibfnamefont {Isaac~L.}\ \bibnamefont
  {Chuang}}, \bibinfo {author} {\bibfnamefont {Luigi}\ \bibnamefont {Frunzio}},
  \bibinfo {author} {\bibfnamefont {Liang}\ \bibnamefont {Jiang}}, \bibinfo
  {author} {\bibfnamefont {S.~M.}\ \bibnamefont {Girvin}}, \ and\ \bibinfo
  {author} {\bibfnamefont {Robert~J.}\ \bibnamefont {Schoelkopf}},\ }\bibfield
  {title} {\enquote {\bibinfo {title} {{Efficient Multiphoton Sampling of
  Molecular Vibronic Spectra on a Superconducting Bosonic Processor}},}\ }\href
  {\doibase 10.1103/physrevx.10.021060} {\bibfield  {journal} {\bibinfo
  {journal} {Phys. Rev. X}\ }\textbf {\bibinfo {volume} {10}},\ \bibinfo
  {pages} {021060} (\bibinfo {year} {2020})}\BibitemShut {NoStop}%
\bibitem [{\citenamefont {Schirmer}(2006)}]{Schirmer2006}%
  \BibitemOpen
  \bibfield  {author} {\bibinfo {author} {\bibfnamefont {Sonia~G}\ \bibnamefont
  {Schirmer}},\ }\bibfield  {title} {\enquote {\bibinfo {title} {{Hamiltonian
  engineering for quantum systems}},}\ }\href
  {http://arxiv.org/abs/quant-ph/0602014} {\bibfield  {journal} {\bibinfo
  {journal} {Lect. Notes Control Inf. Sci.}\ }\textbf {\bibinfo {volume} {366
  LNCIS}},\ \bibinfo {pages} {293} (\bibinfo {year} {2006})}\BibitemShut
  {NoStop}%
\bibitem [{\citenamefont {Goldman}\ and\ \citenamefont
  {Dalibard}(2014)}]{Goldman2014}%
  \BibitemOpen
  \bibfield  {author} {\bibinfo {author} {\bibfnamefont {N.}~\bibnamefont
  {Goldman}}\ and\ \bibinfo {author} {\bibfnamefont {J.}~\bibnamefont
  {Dalibard}},\ }\bibfield  {title} {\enquote {\bibinfo {title} {{Periodically
  driven quantum systems: Effective Hamiltonians and engineered gauge
  fields}},}\ }\href {\doibase 10.1103/PhysRevX.4.031027} {\bibfield  {journal}
  {\bibinfo  {journal} {Phys. Rev. X}\ }\textbf {\bibinfo {volume} {4}},\
  \bibinfo {pages} {031027} (\bibinfo {year} {2014})}\BibitemShut {NoStop}%
\bibitem [{\citenamefont {Krantz}\ \emph {et~al.}(2019)\citenamefont {Krantz},
  \citenamefont {Kjaergaard}, \citenamefont {Yan}, \citenamefont {Orlando},
  \citenamefont {Gustavsson},\ and\ \citenamefont {Oliver}}]{Krantz2019}%
  \BibitemOpen
  \bibfield  {author} {\bibinfo {author} {\bibfnamefont {P.}~\bibnamefont
  {Krantz}}, \bibinfo {author} {\bibfnamefont {M.}~\bibnamefont {Kjaergaard}},
  \bibinfo {author} {\bibfnamefont {F.}~\bibnamefont {Yan}}, \bibinfo {author}
  {\bibfnamefont {T.~P.}\ \bibnamefont {Orlando}}, \bibinfo {author}
  {\bibfnamefont {S.}~\bibnamefont {Gustavsson}}, \ and\ \bibinfo {author}
  {\bibfnamefont {W.~D.}\ \bibnamefont {Oliver}},\ }\bibfield  {title}
  {\enquote {\bibinfo {title} {{A quantum engineer's guide to superconducting
  qubits}},}\ }\href {\doibase 10.1063/1.5089550} {\bibfield  {journal}
  {\bibinfo  {journal} {Appl. Phys. Rev.}\ }\textbf {\bibinfo {volume} {6}},\
  \bibinfo {pages} {021318} (\bibinfo {year} {2019})}\BibitemShut {NoStop}%
\bibitem [{\citenamefont {Haas}\ \emph {et~al.}(2019)\citenamefont {Haas},
  \citenamefont {Puzzuoli}, \citenamefont {Zhang},\ and\ \citenamefont
  {Cory}}]{Haas2019}%
  \BibitemOpen
  \bibfield  {author} {\bibinfo {author} {\bibfnamefont {Holger}\ \bibnamefont
  {Haas}}, \bibinfo {author} {\bibfnamefont {Daniel}\ \bibnamefont {Puzzuoli}},
  \bibinfo {author} {\bibfnamefont {Feihao}\ \bibnamefont {Zhang}}, \ and\
  \bibinfo {author} {\bibfnamefont {David~G}\ \bibnamefont {Cory}},\ }\bibfield
   {title} {\enquote {\bibinfo {title} {{Engineering effective
  Hamiltonians}},}\ }\href {\doibase 10.1088/1367-2630/ab4525} {\bibfield
  {journal} {\bibinfo  {journal} {New J. Phys.}\ }\textbf {\bibinfo {volume}
  {21}},\ \bibinfo {pages} {103011} (\bibinfo {year} {2019})}\BibitemShut
  {NoStop}%
\bibitem [{\citenamefont {Krastanov}\ \emph {et~al.}(2015)\citenamefont
  {Krastanov}, \citenamefont {Albert}, \citenamefont {Shen}, \citenamefont
  {Zou}, \citenamefont {Heeres}, \citenamefont {Vlastakis}, \citenamefont
  {Schoelkopf},\ and\ \citenamefont {Jiang}}]{Krastanov2015}%
  \BibitemOpen
  \bibfield  {author} {\bibinfo {author} {\bibfnamefont {Stefan}\ \bibnamefont
  {Krastanov}}, \bibinfo {author} {\bibfnamefont {Victor~V.}\ \bibnamefont
  {Albert}}, \bibinfo {author} {\bibfnamefont {Chao}\ \bibnamefont {Shen}},
  \bibinfo {author} {\bibfnamefont {C.~L.}\ \bibnamefont {Zou}}, \bibinfo
  {author} {\bibfnamefont {Reinier~W.}\ \bibnamefont {Heeres}}, \bibinfo
  {author} {\bibfnamefont {Brian}\ \bibnamefont {Vlastakis}}, \bibinfo {author}
  {\bibfnamefont {Robert~J.}\ \bibnamefont {Schoelkopf}}, \ and\ \bibinfo
  {author} {\bibfnamefont {Liang}\ \bibnamefont {Jiang}},\ }\bibfield  {title}
  {\enquote {\bibinfo {title} {{Universal control of an oscillator with
  dispersive coupling to a qubit}},}\ }\href {\doibase
  10.1103/PhysRevA.92.040303} {\bibfield  {journal} {\bibinfo  {journal} {Phys.
  Rev. A}\ }\textbf {\bibinfo {volume} {92}},\ \bibinfo {pages} {040303(R)}
  (\bibinfo {year} {2015})}\BibitemShut {NoStop}%
\bibitem [{\citenamefont {Heeres}\ \emph {et~al.}(2015)\citenamefont {Heeres},
  \citenamefont {Vlastakis}, \citenamefont {Holland}, \citenamefont
  {Krastanov}, \citenamefont {Albert}, \citenamefont {Frunzio}, \citenamefont
  {Jiang},\ and\ \citenamefont {Schoelkopf}}]{Heeres2015}%
  \BibitemOpen
  \bibfield  {author} {\bibinfo {author} {\bibfnamefont {Reinier~W}\
  \bibnamefont {Heeres}}, \bibinfo {author} {\bibfnamefont {Brian}\
  \bibnamefont {Vlastakis}}, \bibinfo {author} {\bibfnamefont {Eric}\
  \bibnamefont {Holland}}, \bibinfo {author} {\bibfnamefont {Stefan}\
  \bibnamefont {Krastanov}}, \bibinfo {author} {\bibfnamefont {Victor~V}\
  \bibnamefont {Albert}}, \bibinfo {author} {\bibfnamefont {Luigi}\
  \bibnamefont {Frunzio}}, \bibinfo {author} {\bibfnamefont {Liang}\
  \bibnamefont {Jiang}}, \ and\ \bibinfo {author} {\bibfnamefont {Robert~J}\
  \bibnamefont {Schoelkopf}},\ }\bibfield  {title} {\enquote {\bibinfo {title}
  {{Cavity State Manipulation Using Photon-Number Selective Phase Gates}},}\
  }\href {\doibase 10.1103/PhysRevLett.115.137002} {\bibfield  {journal}
  {\bibinfo  {journal} {Phys. Rev. Lett.}\ }\textbf {\bibinfo {volume} {115}},\
  \bibinfo {pages} {137002} (\bibinfo {year} {2015})}\BibitemShut {NoStop}%
\bibitem [{\citenamefont {Heeres}\ \emph {et~al.}(2017)\citenamefont {Heeres},
  \citenamefont {Reinhold}, \citenamefont {Ofek}, \citenamefont {Frunzio},
  \citenamefont {Jiang}, \citenamefont {Devoret},\ and\ \citenamefont
  {Schoelkopf}}]{Heeres2017}%
  \BibitemOpen
  \bibfield  {author} {\bibinfo {author} {\bibfnamefont {Reinier~W.}\
  \bibnamefont {Heeres}}, \bibinfo {author} {\bibfnamefont {Philip}\
  \bibnamefont {Reinhold}}, \bibinfo {author} {\bibfnamefont {Nissim}\
  \bibnamefont {Ofek}}, \bibinfo {author} {\bibfnamefont {Luigi}\ \bibnamefont
  {Frunzio}}, \bibinfo {author} {\bibfnamefont {Liang}\ \bibnamefont {Jiang}},
  \bibinfo {author} {\bibfnamefont {Michel~H.}\ \bibnamefont {Devoret}}, \ and\
  \bibinfo {author} {\bibfnamefont {Robert~J.}\ \bibnamefont {Schoelkopf}},\
  }\bibfield  {title} {\enquote {\bibinfo {title} {{Implementing a universal
  gate set on a logical qubit encoded in an oscillator}},}\ }\href {\doibase
  10.1038/s41467-017-00045-1} {\bibfield  {journal} {\bibinfo  {journal} {Nat.
  Commun.}\ }\textbf {\bibinfo {volume} {8}},\ \bibinfo {pages} {94} (\bibinfo
  {year} {2017})}\BibitemShut {NoStop}%
\bibitem [{\citenamefont {Gao}\ \emph {et~al.}(2019)\citenamefont {Gao},
  \citenamefont {Lester}, \citenamefont {Chou}, \citenamefont {Frunzio},
  \citenamefont {Devoret}, \citenamefont {Jiang}, \citenamefont {Girvin},\ and\
  \citenamefont {Schoelkopf}}]{Gao2019}%
  \BibitemOpen
  \bibfield  {author} {\bibinfo {author} {\bibfnamefont {Yvonne~Y.}\
  \bibnamefont {Gao}}, \bibinfo {author} {\bibfnamefont {Brian~J.}\
  \bibnamefont {Lester}}, \bibinfo {author} {\bibfnamefont {Kevin~S.}\
  \bibnamefont {Chou}}, \bibinfo {author} {\bibfnamefont {Luigi}\ \bibnamefont
  {Frunzio}}, \bibinfo {author} {\bibfnamefont {Michel~H.}\ \bibnamefont
  {Devoret}}, \bibinfo {author} {\bibfnamefont {Liang}\ \bibnamefont {Jiang}},
  \bibinfo {author} {\bibfnamefont {S.~M.}\ \bibnamefont {Girvin}}, \ and\
  \bibinfo {author} {\bibfnamefont {Robert~J.}\ \bibnamefont {Schoelkopf}},\
  }\bibfield  {title} {\enquote {\bibinfo {title} {{Entanglement of bosonic
  modes through an engineered exchange interaction}},}\ }\href {\doibase
  10.1038/s41586-019-0970-4} {\bibfield  {journal} {\bibinfo  {journal} {Nature
  (London)}\ }\textbf {\bibinfo {volume} {566}},\ \bibinfo {pages} {509}
  (\bibinfo {year} {2019})}\BibitemShut {NoStop}%
\bibitem [{\citenamefont {Boissonneault}\ \emph {et~al.}(2009)\citenamefont
  {Boissonneault}, \citenamefont {Gambetta},\ and\ \citenamefont
  {Blais}}]{Boissonneault2009}%
  \BibitemOpen
  \bibfield  {author} {\bibinfo {author} {\bibfnamefont {Maxime}\ \bibnamefont
  {Boissonneault}}, \bibinfo {author} {\bibfnamefont {J.~M.}\ \bibnamefont
  {Gambetta}}, \ and\ \bibinfo {author} {\bibfnamefont {Alexandre}\
  \bibnamefont {Blais}},\ }\bibfield  {title} {\enquote {\bibinfo {title}
  {{Dispersive regime of circuit QED: Photon-dependent qubit dephasing and
  relaxation rates}},}\ }\href {\doibase 10.1103/PhysRevA.79.013819} {\bibfield
   {journal} {\bibinfo  {journal} {Phys. Rev. A}\ }\textbf {\bibinfo {volume}
  {79}},\ \bibinfo {pages} {013819} (\bibinfo {year} {2009})}\BibitemShut
  {NoStop}%
\bibitem [{\citenamefont {Schuster}\ \emph {et~al.}(2007)\citenamefont
  {Schuster}, \citenamefont {Houck}, \citenamefont {Schreier}, \citenamefont
  {Wallraff}, \citenamefont {Gambetta}, \citenamefont {Blais}, \citenamefont
  {Frunzio}, \citenamefont {Majer}, \citenamefont {Johnson}, \citenamefont
  {Devoret}, \citenamefont {Girvin},\ and\ \citenamefont
  {Schoelkopf}}]{Schuster2007}%
  \BibitemOpen
  \bibfield  {author} {\bibinfo {author} {\bibfnamefont {D~I}\ \bibnamefont
  {Schuster}}, \bibinfo {author} {\bibfnamefont {A~A}\ \bibnamefont {Houck}},
  \bibinfo {author} {\bibfnamefont {J~A}\ \bibnamefont {Schreier}}, \bibinfo
  {author} {\bibfnamefont {A}~\bibnamefont {Wallraff}}, \bibinfo {author}
  {\bibfnamefont {J~M}\ \bibnamefont {Gambetta}}, \bibinfo {author}
  {\bibfnamefont {A}~\bibnamefont {Blais}}, \bibinfo {author} {\bibfnamefont
  {L}~\bibnamefont {Frunzio}}, \bibinfo {author} {\bibfnamefont
  {J}~\bibnamefont {Majer}}, \bibinfo {author} {\bibfnamefont {B}~\bibnamefont
  {Johnson}}, \bibinfo {author} {\bibfnamefont {M~H}\ \bibnamefont {Devoret}},
  \bibinfo {author} {\bibfnamefont {S~M}\ \bibnamefont {Girvin}}, \ and\
  \bibinfo {author} {\bibfnamefont {R~J}\ \bibnamefont {Schoelkopf}},\
  }\bibfield  {title} {\enquote {\bibinfo {title} {{Resolving photon number
  states in a superconducting circuit}},}\ }\href {\doibase
  10.1038/nature05461} {\bibfield  {journal} {\bibinfo  {journal} {Nature
  (London)}\ }\textbf {\bibinfo {volume} {445}},\ \bibinfo {pages} {515}
  (\bibinfo {year} {2007})}\BibitemShut {NoStop}%
\bibitem [{\citenamefont {Gambetta}\ \emph {et~al.}(2006)\citenamefont
  {Gambetta}, \citenamefont {Blais}, \citenamefont {Schuster}, \citenamefont
  {Wallraff}, \citenamefont {Frunzio}, \citenamefont {Majer}, \citenamefont
  {Devoret}, \citenamefont {Girvin},\ and\ \citenamefont
  {Schoelkopf}}]{Gambetta2006}%
  \BibitemOpen
  \bibfield  {author} {\bibinfo {author} {\bibfnamefont {Jay}\ \bibnamefont
  {Gambetta}}, \bibinfo {author} {\bibfnamefont {Alexandre}\ \bibnamefont
  {Blais}}, \bibinfo {author} {\bibfnamefont {D.~I.}\ \bibnamefont {Schuster}},
  \bibinfo {author} {\bibfnamefont {A.}~\bibnamefont {Wallraff}}, \bibinfo
  {author} {\bibfnamefont {L.}~\bibnamefont {Frunzio}}, \bibinfo {author}
  {\bibfnamefont {J.}~\bibnamefont {Majer}}, \bibinfo {author} {\bibfnamefont
  {M.~H.}\ \bibnamefont {Devoret}}, \bibinfo {author} {\bibfnamefont {S.~M.}\
  \bibnamefont {Girvin}}, \ and\ \bibinfo {author} {\bibfnamefont {R.~J.}\
  \bibnamefont {Schoelkopf}},\ }\bibfield  {title} {\enquote {\bibinfo {title}
  {{Qubit-photon interactions in a cavity: Measurement-induced dephasing and
  number splitting}},}\ }\href {\doibase 10.1103/PhysRevA.74.042318} {\bibfield
   {journal} {\bibinfo  {journal} {Phys. Rev. A}\ }\textbf {\bibinfo {volume}
  {74}},\ \bibinfo {pages} {042318} (\bibinfo {year} {2006})}\BibitemShut
  {NoStop}%
\bibitem [{\citenamefont {Johnson}\ \emph {et~al.}(2010)\citenamefont
  {Johnson}, \citenamefont {Reed}, \citenamefont {Houck}, \citenamefont
  {Schuster}, \citenamefont {Bishop}, \citenamefont {Ginossar}, \citenamefont
  {Gambetta}, \citenamefont {Dicarlo}, \citenamefont {Frunzio}, \citenamefont
  {Girvin},\ and\ \citenamefont {Schoelkopf}}]{Johnson2010}%
  \BibitemOpen
  \bibfield  {author} {\bibinfo {author} {\bibfnamefont {B~R}\ \bibnamefont
  {Johnson}}, \bibinfo {author} {\bibfnamefont {M~D}\ \bibnamefont {Reed}},
  \bibinfo {author} {\bibfnamefont {A~A}\ \bibnamefont {Houck}}, \bibinfo
  {author} {\bibfnamefont {D~I}\ \bibnamefont {Schuster}}, \bibinfo {author}
  {\bibfnamefont {Lev~S}\ \bibnamefont {Bishop}}, \bibinfo {author}
  {\bibfnamefont {E}~\bibnamefont {Ginossar}}, \bibinfo {author} {\bibfnamefont
  {J~M}\ \bibnamefont {Gambetta}}, \bibinfo {author} {\bibfnamefont
  {L}~\bibnamefont {Dicarlo}}, \bibinfo {author} {\bibfnamefont
  {L}~\bibnamefont {Frunzio}}, \bibinfo {author} {\bibfnamefont {S~M}\
  \bibnamefont {Girvin}}, \ and\ \bibinfo {author} {\bibfnamefont {R~J}\
  \bibnamefont {Schoelkopf}},\ }\bibfield  {title} {\enquote {\bibinfo {title}
  {{Quantum non-demolition detection of single microwave photons in a
  circuit}},}\ }\href {\doibase 10.1038/nphys1710} {\bibfield  {journal}
  {\bibinfo  {journal} {Nat. Phys.}\ }\textbf {\bibinfo {volume} {6}},\
  \bibinfo {pages} {663} (\bibinfo {year} {2010})}\BibitemShut {NoStop}%
\bibitem [{\citenamefont {Rahav}\ \emph {et~al.}(2003)\citenamefont {Rahav},
  \citenamefont {Gilary},\ and\ \citenamefont {Fishman}}]{Rahav2003}%
  \BibitemOpen
  \bibfield  {author} {\bibinfo {author} {\bibfnamefont {Saar}\ \bibnamefont
  {Rahav}}, \bibinfo {author} {\bibfnamefont {Ido}\ \bibnamefont {Gilary}}, \
  and\ \bibinfo {author} {\bibfnamefont {Shmuel}\ \bibnamefont {Fishman}},\
  }\bibfield  {title} {\enquote {\bibinfo {title} {{Effective Hamiltonians for
  periodically driven systems}},}\ }\href {\doibase 10.1103/PhysRevA.68.013820}
  {\bibfield  {journal} {\bibinfo  {journal} {Phys. Rev. A}\ }\textbf {\bibinfo
  {volume} {68}},\ \bibinfo {pages} {013820} (\bibinfo {year}
  {2003})}\BibitemShut {NoStop}%
\bibitem [{\citenamefont {Reagor}\ \emph {et~al.}(2013)\citenamefont {Reagor},
  \citenamefont {Paik}, \citenamefont {Catelani}, \citenamefont {Sun},
  \citenamefont {Axline}, \citenamefont {Holland}, \citenamefont {Pop},
  \citenamefont {Masluk}, \citenamefont {Brecht}, \citenamefont {Frunzio},
  \citenamefont {Devoret}, \citenamefont {Glazman},\ and\ \citenamefont
  {Schoelkopf}}]{Reagor2013}%
  \BibitemOpen
  \bibfield  {author} {\bibinfo {author} {\bibfnamefont {Matthew}\ \bibnamefont
  {Reagor}}, \bibinfo {author} {\bibfnamefont {Hanhee}\ \bibnamefont {Paik}},
  \bibinfo {author} {\bibfnamefont {Gianluigi}\ \bibnamefont {Catelani}},
  \bibinfo {author} {\bibfnamefont {Luyan}\ \bibnamefont {Sun}}, \bibinfo
  {author} {\bibfnamefont {Christopher}\ \bibnamefont {Axline}}, \bibinfo
  {author} {\bibfnamefont {Eric}\ \bibnamefont {Holland}}, \bibinfo {author}
  {\bibfnamefont {Ioan~M.}\ \bibnamefont {Pop}}, \bibinfo {author}
  {\bibfnamefont {Nicholas~A.}\ \bibnamefont {Masluk}}, \bibinfo {author}
  {\bibfnamefont {Teresa}\ \bibnamefont {Brecht}}, \bibinfo {author}
  {\bibfnamefont {Luigi}\ \bibnamefont {Frunzio}}, \bibinfo {author}
  {\bibfnamefont {Michel~H.}\ \bibnamefont {Devoret}}, \bibinfo {author}
  {\bibfnamefont {Leonid}\ \bibnamefont {Glazman}}, \ and\ \bibinfo {author}
  {\bibfnamefont {Robert~J.}\ \bibnamefont {Schoelkopf}},\ }\bibfield  {title}
  {\enquote {\bibinfo {title} {{Reaching 10 ms single photon lifetimes for
  superconducting aluminum cavities}},}\ }\href {\doibase 10.1063/1.4807015}
  {\bibfield  {journal} {\bibinfo  {journal} {Appl. Phys. Lett.}\ }\textbf
  {\bibinfo {volume} {102}},\ \bibinfo {pages} {192604} (\bibinfo {year}
  {2013})}\BibitemShut {NoStop}%
\bibitem [{\citenamefont {Lebreuilly}\ \emph {et~al.}(2021)\citenamefont
  {Lebreuilly}, \citenamefont {Noh}, \citenamefont {Wang}, \citenamefont
  {Girvin},\ and\ \citenamefont {Jiang}}]{Lebreuilly2021}%
  \BibitemOpen
  \bibfield  {author} {\bibinfo {author} {\bibfnamefont {Jos{\'{e}}}\
  \bibnamefont {Lebreuilly}}, \bibinfo {author} {\bibfnamefont {Kyungjoo}\
  \bibnamefont {Noh}}, \bibinfo {author} {\bibfnamefont {Chiao-Hsuan}\
  \bibnamefont {Wang}}, \bibinfo {author} {\bibfnamefont {Steven~M.}\
  \bibnamefont {Girvin}}, \ and\ \bibinfo {author} {\bibfnamefont {Liang}\
  \bibnamefont {Jiang}},\ }\bibfield  {title} {\enquote {\bibinfo {title}
  {{Autonomous quantum error correction and quantum computation}},}\ }\href
  {http://arxiv.org/abs/2103.05007} {\  (\bibinfo {year} {2021})},\ \Eprint
  {http://arxiv.org/abs/2103.05007} {arXiv:2103.05007} \BibitemShut {NoStop}%
\bibitem [{\citenamefont {Nielsen}\ and\ \citenamefont
  {Chuang}(2010)}]{Nielsen2010}%
  \BibitemOpen
  \bibfield  {author} {\bibinfo {author} {\bibfnamefont {Michael~A}\
  \bibnamefont {Nielsen}}\ and\ \bibinfo {author} {\bibfnamefont {Issac}\
  \bibnamefont {Chuang}},\ }\href@noop {} {\emph {\bibinfo {title} {{Quantum
  Computation and Quantum Information: 10th Anniversary Edition}}}}\ (\bibinfo
  {publisher} {Cambridge University Press, Cambridge, UK},\ \bibinfo {year}
  {2010})\BibitemShut {NoStop}%
\bibitem [{\citenamefont {Vy}\ \emph {et~al.}(2013)\citenamefont {Vy},
  \citenamefont {Wang},\ and\ \citenamefont {Jacobs}}]{Vy2013}%
  \BibitemOpen
  \bibfield  {author} {\bibinfo {author} {\bibfnamefont {Os}~\bibnamefont
  {Vy}}, \bibinfo {author} {\bibfnamefont {Xiaoting}\ \bibnamefont {Wang}}, \
  and\ \bibinfo {author} {\bibfnamefont {Kurt}\ \bibnamefont {Jacobs}},\
  }\bibfield  {title} {\enquote {\bibinfo {title} {{Error-transparent
  evolution: The ability of multi-body interactions to bypass decoherence}},}\
  }\href {\doibase 10.1088/1367-2630/15/5/053002} {\bibfield  {journal}
  {\bibinfo  {journal} {New J. Phys.}\ }\textbf {\bibinfo {volume} {15}},\
  \bibinfo {pages} {053002} (\bibinfo {year} {2013})}\BibitemShut {NoStop}%
\bibitem [{\citenamefont {Kapit}(2018)}]{Kapit2018}%
  \BibitemOpen
  \bibfield  {author} {\bibinfo {author} {\bibfnamefont {Eliot}\ \bibnamefont
  {Kapit}},\ }\bibfield  {title} {\enquote {\bibinfo {title}
  {{Error-Transparent Quantum Gates for Small Logical Qubit Architectures}},}\
  }\href {\doibase 10.1103/PhysRevLett.120.050503} {\bibfield  {journal}
  {\bibinfo  {journal} {Phys. Rev. Lett.}\ }\textbf {\bibinfo {volume} {120}},\
  \bibinfo {pages} {050503} (\bibinfo {year} {2018})}\BibitemShut {NoStop}%
\bibitem [{\citenamefont {Rosenblum}\ \emph {et~al.}(2018)\citenamefont
  {Rosenblum}, \citenamefont {Reinhold}, \citenamefont {Mirrahimi},
  \citenamefont {Jiang}, \citenamefont {Frunzio},\ and\ \citenamefont
  {Schoelkopf}}]{Rosenblum2018}%
  \BibitemOpen
  \bibfield  {author} {\bibinfo {author} {\bibfnamefont {S.}~\bibnamefont
  {Rosenblum}}, \bibinfo {author} {\bibfnamefont {P.}~\bibnamefont {Reinhold}},
  \bibinfo {author} {\bibfnamefont {M.}~\bibnamefont {Mirrahimi}}, \bibinfo
  {author} {\bibfnamefont {Liang}\ \bibnamefont {Jiang}}, \bibinfo {author}
  {\bibfnamefont {L.}~\bibnamefont {Frunzio}}, \ and\ \bibinfo {author}
  {\bibfnamefont {R.~J.}\ \bibnamefont {Schoelkopf}},\ }\bibfield  {title}
  {\enquote {\bibinfo {title} {{Fault-tolerant detection of a quantum
  error}},}\ }\href {\doibase 10.1126/science.aat3996} {\bibfield  {journal}
  {\bibinfo  {journal} {Science}\ }\textbf {\bibinfo {volume} {361}},\ \bibinfo
  {pages} {266} (\bibinfo {year} {2018})}\BibitemShut {NoStop}%
\bibitem [{\citenamefont {Ma}\ \emph {et~al.}(2020{\natexlab{a}})\citenamefont
  {Ma}, \citenamefont {Xu}, \citenamefont {Mu}, \citenamefont {Cai},
  \citenamefont {Hu}, \citenamefont {Wang}, \citenamefont {Pan}, \citenamefont
  {Wang}, \citenamefont {Song}, \citenamefont {Zou},\ and\ \citenamefont
  {Sun}}]{Sun2020}%
  \BibitemOpen
  \bibfield  {author} {\bibinfo {author} {\bibfnamefont {Y.}~\bibnamefont
  {Ma}}, \bibinfo {author} {\bibfnamefont {Y.}~\bibnamefont {Xu}}, \bibinfo
  {author} {\bibfnamefont {X.}~\bibnamefont {Mu}}, \bibinfo {author}
  {\bibfnamefont {W.}~\bibnamefont {Cai}}, \bibinfo {author} {\bibfnamefont
  {L.}~\bibnamefont {Hu}}, \bibinfo {author} {\bibfnamefont {W.}~\bibnamefont
  {Wang}}, \bibinfo {author} {\bibfnamefont {X.}~\bibnamefont {Pan}}, \bibinfo
  {author} {\bibfnamefont {H.}~\bibnamefont {Wang}}, \bibinfo {author}
  {\bibfnamefont {Y.~P.}\ \bibnamefont {Song}}, \bibinfo {author}
  {\bibfnamefont {C.~L.}\ \bibnamefont {Zou}}, \ and\ \bibinfo {author}
  {\bibfnamefont {L.}~\bibnamefont {Sun}},\ }\bibfield  {title} {\enquote
  {\bibinfo {title} {{Error-transparent operations on a logical qubit protected
  by quantum error correction}},}\ }\href {\doibase 10.1038/s41567-020-0893-x}
  {\bibfield  {journal} {\bibinfo  {journal} {Nat. Phys.}\ }\textbf {\bibinfo
  {volume} {16}},\ \bibinfo {pages} {827} (\bibinfo {year}
  {2020}{\natexlab{a}})}\BibitemShut {NoStop}%
\bibitem [{\citenamefont {Lihm}\ \emph {et~al.}(2018)\citenamefont {Lihm},
  \citenamefont {Noh},\ and\ \citenamefont {Fischer}}]{Lihm2018}%
  \BibitemOpen
  \bibfield  {author} {\bibinfo {author} {\bibfnamefont {J.~M.}\ \bibnamefont
  {Lihm}}, \bibinfo {author} {\bibfnamefont {Kyungjoo}\ \bibnamefont {Noh}}, \
  and\ \bibinfo {author} {\bibfnamefont {U.~R.}\ \bibnamefont {Fischer}},\
  }\bibfield  {title} {\enquote {\bibinfo {title} {{Implementation-independent
  sufficient condition of the Knill-Laflamme type for the autonomous protection
  of logical qudits by strong engineered dissipation}},}\ }\href {\doibase
  10.1103/PhysRevA.98.012317} {\bibfield  {journal} {\bibinfo  {journal} {Phys.
  Rev. A}\ }\textbf {\bibinfo {volume} {98}},\ \bibinfo {pages} {012317}
  (\bibinfo {year} {2018})}\BibitemShut {NoStop}%
\bibitem [{\citenamefont {Koch}\ \emph {et~al.}(2007)\citenamefont {Koch},
  \citenamefont {Yu}, \citenamefont {Gambetta}, \citenamefont {Houck},
  \citenamefont {Schuster}, \citenamefont {Majer}, \citenamefont {Blais},
  \citenamefont {Devoret}, \citenamefont {Girvin},\ and\ \citenamefont
  {Schoelkopf}}]{Koch2007}%
  \BibitemOpen
  \bibfield  {author} {\bibinfo {author} {\bibfnamefont {J.}~\bibnamefont
  {Koch}}, \bibinfo {author} {\bibfnamefont {T.~M.}\ \bibnamefont {Yu}},
  \bibinfo {author} {\bibfnamefont {J.}~\bibnamefont {Gambetta}}, \bibinfo
  {author} {\bibfnamefont {A.~A.}\ \bibnamefont {Houck}}, \bibinfo {author}
  {\bibfnamefont {D.~I.}\ \bibnamefont {Schuster}}, \bibinfo {author}
  {\bibfnamefont {J.}~\bibnamefont {Majer}}, \bibinfo {author} {\bibfnamefont
  {A.}~\bibnamefont {Blais}}, \bibinfo {author} {\bibfnamefont {M.~H.}\
  \bibnamefont {Devoret}}, \bibinfo {author} {\bibfnamefont {S.~M.}\
  \bibnamefont {Girvin}}, \ and\ \bibinfo {author} {\bibfnamefont {R.~J.}\
  \bibnamefont {Schoelkopf}},\ }\bibfield  {title} {\enquote {\bibinfo {title}
  {{Charge-insensitive qubit design derived from the Cooper pair box}},}\
  }\href {\doibase 10.1103/PhysRevA.76.042319} {\bibfield  {journal} {\bibinfo
  {journal} {Phys. Rev. A}\ }\textbf {\bibinfo {volume} {76}},\ \bibinfo
  {pages} {042319} (\bibinfo {year} {2007})}\BibitemShut {NoStop}%
\bibitem [{\citenamefont {Kirchmair}\ \emph {et~al.}(2013)\citenamefont
  {Kirchmair}, \citenamefont {Vlastakis}, \citenamefont {Leghtas},
  \citenamefont {Nigg}, \citenamefont {Paik}, \citenamefont {Ginossar},
  \citenamefont {Mirrahimi}, \citenamefont {Frunzio}, \citenamefont {Girvin},\
  and\ \citenamefont {Schoelkopf}}]{Kirchmair2013}%
  \BibitemOpen
  \bibfield  {author} {\bibinfo {author} {\bibfnamefont {Gerhard}\ \bibnamefont
  {Kirchmair}}, \bibinfo {author} {\bibfnamefont {Brian}\ \bibnamefont
  {Vlastakis}}, \bibinfo {author} {\bibfnamefont {Zaki}\ \bibnamefont
  {Leghtas}}, \bibinfo {author} {\bibfnamefont {Simon~E.}\ \bibnamefont
  {Nigg}}, \bibinfo {author} {\bibfnamefont {Hanhee}\ \bibnamefont {Paik}},
  \bibinfo {author} {\bibfnamefont {Eran}\ \bibnamefont {Ginossar}}, \bibinfo
  {author} {\bibfnamefont {Mazyar}\ \bibnamefont {Mirrahimi}}, \bibinfo
  {author} {\bibfnamefont {Luigi}\ \bibnamefont {Frunzio}}, \bibinfo {author}
  {\bibfnamefont {S.~M.}\ \bibnamefont {Girvin}}, \ and\ \bibinfo {author}
  {\bibfnamefont {R.~J.}\ \bibnamefont {Schoelkopf}},\ }\bibfield  {title}
  {\enquote {\bibinfo {title} {{Observation of quantum state collapse and
  revival due to the single-photon Kerr effect}},}\ }\href {\doibase
  10.1038/nature11902} {\bibfield  {journal} {\bibinfo  {journal} {Nature
  (London)}\ }\textbf {\bibinfo {volume} {495}},\ \bibinfo {pages} {205}
  (\bibinfo {year} {2013})}\BibitemShut {NoStop}%
\bibitem [{\citenamefont {Xu}\ \emph {et~al.}(2020)\citenamefont {Xu},
  \citenamefont {Ma}, \citenamefont {Cai}, \citenamefont {Mu}, \citenamefont
  {Dai}, \citenamefont {Wang}, \citenamefont {Hu}, \citenamefont {Li},
  \citenamefont {Han}, \citenamefont {Wang}, \citenamefont {Song},
  \citenamefont {Yang}, \citenamefont {Zheng},\ and\ \citenamefont
  {Sun}}]{Xu2020}%
  \BibitemOpen
  \bibfield  {author} {\bibinfo {author} {\bibfnamefont {Y.}~\bibnamefont
  {Xu}}, \bibinfo {author} {\bibfnamefont {Y.}~\bibnamefont {Ma}}, \bibinfo
  {author} {\bibfnamefont {W.}~\bibnamefont {Cai}}, \bibinfo {author}
  {\bibfnamefont {X.}~\bibnamefont {Mu}}, \bibinfo {author} {\bibfnamefont
  {W.}~\bibnamefont {Dai}}, \bibinfo {author} {\bibfnamefont {W.}~\bibnamefont
  {Wang}}, \bibinfo {author} {\bibfnamefont {L.}~\bibnamefont {Hu}}, \bibinfo
  {author} {\bibfnamefont {X.}~\bibnamefont {Li}}, \bibinfo {author}
  {\bibfnamefont {J.}~\bibnamefont {Han}}, \bibinfo {author} {\bibfnamefont
  {H.}~\bibnamefont {Wang}}, \bibinfo {author} {\bibfnamefont {Y.~P.}\
  \bibnamefont {Song}}, \bibinfo {author} {\bibfnamefont {Zhen~Biao}\
  \bibnamefont {Yang}}, \bibinfo {author} {\bibfnamefont {Shi~Biao}\
  \bibnamefont {Zheng}}, \ and\ \bibinfo {author} {\bibfnamefont
  {L.}~\bibnamefont {Sun}},\ }\bibfield  {title} {\enquote {\bibinfo {title}
  {{Demonstration of Controlled-Phase Gates between Two Error-Correctable
  Photonic Qubits}},}\ }\href {\doibase 10.1103/PhysRevLett.124.120501}
  {\bibfield  {journal} {\bibinfo  {journal} {Phys. Rev. Lett.}\ }\textbf
  {\bibinfo {volume} {124}},\ \bibinfo {pages} {120501} (\bibinfo {year}
  {2020})}\BibitemShut {NoStop}%
\bibitem [{\citenamefont {Sletten}\ \emph {et~al.}(2019)\citenamefont
  {Sletten}, \citenamefont {Moores}, \citenamefont {Viennot},\ and\
  \citenamefont {Lehnert}}]{Sletten2019}%
  \BibitemOpen
  \bibfield  {author} {\bibinfo {author} {\bibfnamefont {L.~R.}\ \bibnamefont
  {Sletten}}, \bibinfo {author} {\bibfnamefont {B.~A.}\ \bibnamefont {Moores}},
  \bibinfo {author} {\bibfnamefont {J~.J.}\ \bibnamefont {Viennot}}, \ and\
  \bibinfo {author} {\bibfnamefont {K.~W.}\ \bibnamefont {Lehnert}},\
  }\bibfield  {title} {\enquote {\bibinfo {title} {{Resolving Phonon Fock
  States in a Multimode Cavity with a Double-Slit Qubit}},}\ }\href {\doibase
  10.1103/PhysRevX.9.021056} {\bibfield  {journal} {\bibinfo  {journal} {Phys.
  Rev. X}\ }\textbf {\bibinfo {volume} {9}},\ \bibinfo {pages} {021056}
  (\bibinfo {year} {2019})}\BibitemShut {NoStop}%
\bibitem [{\citenamefont {Arrangoiz-Arriola}\ \emph {et~al.}(2019)\citenamefont
  {Arrangoiz-Arriola}, \citenamefont {Wollack}, \citenamefont {Wang},
  \citenamefont {Pechal}, \citenamefont {Jiang}, \citenamefont {McKenna},
  \citenamefont {Witmer}, \citenamefont {{Van Laer}},\ and\ \citenamefont
  {Safavi-Naeini}}]{Arrangoiz2019}%
  \BibitemOpen
  \bibfield  {author} {\bibinfo {author} {\bibfnamefont {Patricio}\
  \bibnamefont {Arrangoiz-Arriola}}, \bibinfo {author} {\bibfnamefont
  {E.~Alex}\ \bibnamefont {Wollack}}, \bibinfo {author} {\bibfnamefont
  {Zhaoyou}\ \bibnamefont {Wang}}, \bibinfo {author} {\bibfnamefont {Marek}\
  \bibnamefont {Pechal}}, \bibinfo {author} {\bibfnamefont {Wentao}\
  \bibnamefont {Jiang}}, \bibinfo {author} {\bibfnamefont {Timothy~P.}\
  \bibnamefont {McKenna}}, \bibinfo {author} {\bibfnamefont {Jeremy~D.}\
  \bibnamefont {Witmer}}, \bibinfo {author} {\bibfnamefont {Rapha{\"{e}}l}\
  \bibnamefont {{Van Laer}}}, \ and\ \bibinfo {author} {\bibfnamefont
  {Amir~H.}\ \bibnamefont {Safavi-Naeini}},\ }\bibfield  {title} {\enquote
  {\bibinfo {title} {{Resolving the energy levels of a nanomechanical
  oscillator}},}\ }\href {\doibase 10.1038/s41586-019-1386-x} {\bibfield
  {journal} {\bibinfo  {journal} {Nature (London)}\ }\textbf {\bibinfo {volume}
  {571}},\ \bibinfo {pages} {537} (\bibinfo {year} {2019})}\BibitemShut
  {NoStop}%
\bibitem [{\citenamefont {Reinhold}\ \emph {et~al.}(2020)\citenamefont
  {Reinhold}, \citenamefont {Rosenblum}, \citenamefont {Ma}, \citenamefont
  {Frunzio}, \citenamefont {Jiang},\ and\ \citenamefont
  {Schoelkopf}}]{Reinhold2020}%
  \BibitemOpen
  \bibfield  {author} {\bibinfo {author} {\bibfnamefont {Philip}\ \bibnamefont
  {Reinhold}}, \bibinfo {author} {\bibfnamefont {Serge}\ \bibnamefont
  {Rosenblum}}, \bibinfo {author} {\bibfnamefont {Wen~Long}\ \bibnamefont
  {Ma}}, \bibinfo {author} {\bibfnamefont {Luigi}\ \bibnamefont {Frunzio}},
  \bibinfo {author} {\bibfnamefont {Liang}\ \bibnamefont {Jiang}}, \ and\
  \bibinfo {author} {\bibfnamefont {Robert~J.}\ \bibnamefont {Schoelkopf}},\
  }\bibfield  {title} {\enquote {\bibinfo {title} {{Error-corrected gates on an
  encoded qubit}},}\ }\href {\doibase 10.1038/s41567-020-0931-8} {\bibfield
  {journal} {\bibinfo  {journal} {Nat. Phys.}\ }\textbf {\bibinfo {volume}
  {16}},\ \bibinfo {pages} {822} (\bibinfo {year} {2020})}\BibitemShut
  {NoStop}%
\bibitem [{\citenamefont {Ma}\ \emph {et~al.}(2020{\natexlab{b}})\citenamefont
  {Ma}, \citenamefont {Zhang}, \citenamefont {Wong}, \citenamefont {Noh},
  \citenamefont {Rosenblum}, \citenamefont {Reinhold}, \citenamefont
  {Schoelkopf},\ and\ \citenamefont {Jiang}}]{Ma2020}%
  \BibitemOpen
  \bibfield  {author} {\bibinfo {author} {\bibfnamefont {Wen~Long}\
  \bibnamefont {Ma}}, \bibinfo {author} {\bibfnamefont {Mengzhen}\ \bibnamefont
  {Zhang}}, \bibinfo {author} {\bibfnamefont {Yat}\ \bibnamefont {Wong}},
  \bibinfo {author} {\bibfnamefont {Kyungjoo}\ \bibnamefont {Noh}}, \bibinfo
  {author} {\bibfnamefont {Serge}\ \bibnamefont {Rosenblum}}, \bibinfo {author}
  {\bibfnamefont {Philip}\ \bibnamefont {Reinhold}}, \bibinfo {author}
  {\bibfnamefont {Robert~J.}\ \bibnamefont {Schoelkopf}}, \ and\ \bibinfo
  {author} {\bibfnamefont {Liang}\ \bibnamefont {Jiang}},\ }\bibfield  {title}
  {\enquote {\bibinfo {title} {{Path-Independent Quantum Gates with Noisy
  Ancilla}},}\ }\href {\doibase 10.1103/PHYSREVLETT.125.110503} {\bibfield
  {journal} {\bibinfo  {journal} {Phys. Rev. Lett.}\ }\textbf {\bibinfo
  {volume} {125}},\ \bibinfo {pages} {110503} (\bibinfo {year}
  {2020}{\natexlab{b}})}\BibitemShut {NoStop}%
\bibitem [{\citenamefont {Kapit}(2016)}]{Kapit2016}%
  \BibitemOpen
  \bibfield  {author} {\bibinfo {author} {\bibfnamefont {Eliot}\ \bibnamefont
  {Kapit}},\ }\bibfield  {title} {\enquote {\bibinfo {title}
  {{Hardware-efficient and fully autonomous quantum error correction in
  superconducting circuits}},}\ }\href {\doibase
  10.1103/PhysRevLett.116.150501} {\bibfield  {journal} {\bibinfo  {journal}
  {Phys. Rev. Lett.}\ }\textbf {\bibinfo {volume} {116}},\ \bibinfo {pages}
  {150501} (\bibinfo {year} {2016})}\BibitemShut {NoStop}%
\bibitem [{\citenamefont {Dutta}\ \emph {et~al.}(2015)\citenamefont {Dutta},
  \citenamefont {Gajda}, \citenamefont {Hauke}, \citenamefont {Lewenstein},
  \citenamefont {L{\"{u}}hmann}, \citenamefont {Malomed}, \citenamefont
  {Sowi{\'{n}}ski},\ and\ \citenamefont {Zakrzewski}}]{Dutta2015}%
  \BibitemOpen
  \bibfield  {author} {\bibinfo {author} {\bibfnamefont {Omjyoti}\ \bibnamefont
  {Dutta}}, \bibinfo {author} {\bibfnamefont {Mariusz}\ \bibnamefont {Gajda}},
  \bibinfo {author} {\bibfnamefont {Philipp}\ \bibnamefont {Hauke}}, \bibinfo
  {author} {\bibfnamefont {Maciej}\ \bibnamefont {Lewenstein}}, \bibinfo
  {author} {\bibfnamefont {Dirk-S{\"{o}}ren}\ \bibnamefont {L{\"{u}}hmann}},
  \bibinfo {author} {\bibfnamefont {Boris~A}\ \bibnamefont {Malomed}}, \bibinfo
  {author} {\bibfnamefont {Tomasz}\ \bibnamefont {Sowi{\'{n}}ski}}, \ and\
  \bibinfo {author} {\bibfnamefont {Jakub}\ \bibnamefont {Zakrzewski}},\
  }\bibfield  {title} {\enquote {\bibinfo {title} {{Non-standard Hubbard models
  in optical lattices: a review}},}\ }\href {\doibase
  10.1088/0034-4885/78/6/066001} {\bibfield  {journal} {\bibinfo  {journal}
  {Reports Prog. Phys.}\ }\textbf {\bibinfo {volume} {78}},\ \bibinfo {pages}
  {066001} (\bibinfo {year} {2015})}\BibitemShut {NoStop}%
\bibitem [{\citenamefont {Dai}\ \emph {et~al.}(2016)\citenamefont {Dai},
  \citenamefont {Shi},\ and\ \citenamefont {Yi}}]{Dai2016}%
  \BibitemOpen
  \bibfield  {author} {\bibinfo {author} {\bibfnamefont {C.~M.}\ \bibnamefont
  {Dai}}, \bibinfo {author} {\bibfnamefont {Z.~C.}\ \bibnamefont {Shi}}, \ and\
  \bibinfo {author} {\bibfnamefont {X.~X.}\ \bibnamefont {Yi}},\ }\bibfield
  {title} {\enquote {\bibinfo {title} {{Floquet theorem with open systems and
  its applications}},}\ }\href {\doibase 10.1103/PhysRevA.93.032121} {\bibfield
   {journal} {\bibinfo  {journal} {Phys. Rev. A}\ }\textbf {\bibinfo {volume}
  {93}},\ \bibinfo {pages} {032121} (\bibinfo {year} {2016})}\BibitemShut
  {NoStop}%
\bibitem [{\citenamefont {Scopa}\ \emph {et~al.}(2019)\citenamefont {Scopa},
  \citenamefont {Landi}, \citenamefont {Hammoumi},\ and\ \citenamefont
  {Karevski}}]{Scopa2019}%
  \BibitemOpen
  \bibfield  {author} {\bibinfo {author} {\bibfnamefont {Stefano}\ \bibnamefont
  {Scopa}}, \bibinfo {author} {\bibfnamefont {Gabriel~T}\ \bibnamefont
  {Landi}}, \bibinfo {author} {\bibfnamefont {Adam}\ \bibnamefont {Hammoumi}},
  \ and\ \bibinfo {author} {\bibfnamefont {Dragi}\ \bibnamefont {Karevski}},\
  }\bibfield  {title} {\enquote {\bibinfo {title} {{Exact solution of
  time-dependent Lindblad equations with closed algebras}},}\ }\href {\doibase
  10.1103/PhysRevA.99.022105} {\bibfield  {journal} {\bibinfo  {journal} {Phys.
  Rev. A}\ }\textbf {\bibinfo {volume} {99}},\ \bibinfo {pages} {022105}
  (\bibinfo {year} {2019})}\BibitemShut {NoStop}%
\end{thebibliography}%
\end{document}